\RequirePackage{fix-cm}
\documentclass[smallcondensed, twocolumn]{svjour2}          
\smartqed  
\usepackage{graphicx}
\usepackage{dcolumn}
\usepackage{bm}
\usepackage{amssymb}   
\usepackage{subfigure}
\usepackage{tikz}
\usepackage{epstopdf}
\usepackage{epsfig}
\usepackage{upgreek}
\usepackage{url}
\usepackage{cite} 
\usepackage[english]{babel}
\usepackage{textcomp}
\usepackage{amsmath}

\usepackage[colorlinks=true, allcolors=blue]{hyperref}

\usepackage[sort&compress, numbers]{natbib}

\begin{document}\sloppy

\title{Structural transitions in jammed asymmetric bidisperse granular packings}


\author{Juan C. Petit \and Matthias Sperl}
\institute{Juan C. Petit \and Matthias Sperl 
\at German Aerospace Center, \\
Institute of Materials Physics in Space \\
Linder H\"ohe, 51147 K\"oln, \\
\email{Juan.Petit@dlr.de, Matthias.Sperl@dlr.de}
}


\date{\today}

\maketitle

\begin{abstract}

We study the local structural changes along the jamming transitions in asymmetric 
bidisperse granu\-lar packings. The local structure of the packing is assessed by 
the contact orientational order, $\tilde{Q}_{\ell}$, that quantifies the 
contribution of each contact configuration (Large-Large, Small-Small, Large-Small, Small-Large)
in the jammed structure. The partial values of $\tilde{Q}_{\ell}$ are calculated with respect to 
known ordered lattices that are fixed by the size ratio, $\delta$, of the particles.
We find that the packing undergoes a structural transition at $\phi_J$,  
manifested by a sudden jump in the partial $\tilde{Q}_{\ell}$. Each contact configuration 
contributes to the jammed structure in a different way, changing with $\delta$ and 
concentration of small particles, $X_{\mathrm{S}}$. 
The results show not only that the packing undergoes a structural change upon jamming, 
but also that bidisperse packings exhibit local HCP and FCC structures 
also found in monodisperse packings. This suggests that the jammed
structure of bidisperse systems is inherently endowed with local structural order. These 
results are relevant in understanding how the arrangement of particles determines the 
strength of bidisperse granular packings.

\keywords{Jamming transition \and Structure \and Amorphous \and Extreme size ratio \and Bidisperse packings}

\end{abstract}


\section{Introduction}
\label{SecI}

The jamming transition in granular packings has been studied for years, with 
much attention paid to monodisperse packings since this is the simplest case 
\cite{donev2004jamming, ohern2003jamming, silbert2009long}. 
Such a transition is defined when a set of non-contacting spheres come into contact 
collectively to form a rigid structure. For a monodisperse packing, the jamming 
transition occurs at a jamming density around $\phi_J \approx 0.64$ in 3D. 
Considering a second particle size in the packing with a size 
ratio of $\delta = r_{\mathrm{S}}/r_{\mathrm{L}} = 0.71$ and the same number of 
large and small particles (50:50 mixture), $\phi_J$ increases slightly compared 
to the monodisperse case \cite{majmudar2007jamming, ohern2003jamming, morse2016geometric}. 
On the other hand, varying the concentration of small particles, $X_{\mathrm{S}}$, 
and size ratio, $\delta$, studies have shown a richer jamming diagram for 
bidisperse packings than the monodisperse and even the 50:50 mixture with $\delta = 0.71$
\cite{prasad2017subjamming,hopkins2011phase,frank2014disordered, biazzo2009theory, 
pillitteri2019jamming, kumar2016tuning, hara2021phase, koeze2016mapping, petit2020additional, 
petit2022bulk}. In this case, $\phi_J$ shows a maximum value at a given $X_{\mathrm{S}}$ that 
increases as $\delta$ decreases. Similar results have also been obtained 
in dense suspension where the viscosity was found to be minimal at a specific value of $X_{\mathrm{S}}$ 
while the jamming density was found to be maximal for the same $X_{\mathrm{S}}$ 
\cite{farris1968prediction, shapiro1992random, pednekar2018bidisperse, he2001viscosity, gondret1997dynamic}. 
Such a minimal value of the viscosity also decreases whereas the maximal value 
of the jamming density increases when $\delta$ decreases.

Recently, it was shown that there are critical $\delta$ and $X_{\mathrm{S}}$ values 
below which the jamming structure of a bidisperse system consists only of large 
particles, while most small particles remain without contacts \cite{prasad2017subjamming, 
hara2021phase, petit2020additional, petit2022bulk}. This suggests that the jammed structure can be regarded as
a monodisperse rather than a bidisperse packing, although the packing itself can still be 
modified by the presence of small particles at higher compression. This finding has led to reconsider the 
jamming transition diagram for bidisperse packings to provide a more general overview. In
 a recent paper, it is shown that at low $\delta$ and low $X_{\mathrm{S}}$, small 
 particles can be jammed by compressing beyond the 
$\phi_J$ formed by large particles \cite{petit2020additional}. Such $\phi_J$ is extracted 
at a packing fraction where the fraction of large particles, $n_{\mathrm{L}}$, contributing 
to the jammed structure exhibits a jump, defined by a sharp but finite value similar
to that observed in the mean contact number, see Ref.~\cite{petit2020additional}. A 
separate jump in the fraction of small particles, 
$n_{\mathrm{S}}$, occurs at $\phi > \phi_J$, which was interpreted as the jammed 
transition of the small particles. This result led to the idea that a bidisperse 
packing in compression has two jamming transitions at low $\delta$ and low $X_{\mathrm{S}}$.
The first jamming transition is driven by the 
jamming of predominantly large particles and the second transition small particles are 
jammed together with large ones. This second transition was shown for $\delta \leq 0.22$ 
to be an additional line extending towards higher packing densities as $X_{\mathrm{S}}$ is 
reduced, generalizing the jamming diagram of bidisperse packings.

The evolution of the jammed structure in a range of packing fractions has been well 
studied in monodisperse hard-sphere packings \citep{clarke1993structural, 
klumov2011structural, klumov2014structural, hanifpour2015structural}, showing that 
upon compression there is a structural transition from disordered to an ordered local 
structure. All works report the development of local Hexagonal Close-Packed (HCP) and 
Face-Centred Cubic (FCC) lattices as the system becomes denser. The evolution of the 
jammed structure in bidisperse packings has not been explored in detail. It is not 
clear how each configuration type; Large-Large (LL), Small-Small (SS), Large-Small (LS), 
and Small-Large (SL), contribute to the development of the jammed structure when $\delta$ 
and $X_{\mathrm{S}}$ are varied. The way each particle size is packed in the system is 
important to understand the transition to jamming and also how such structures can lead 
to different structural properties. In this work, we investigate the structural evolution 
of jammed bidisperse packings along the first and second jamming transition lines recently 
reported. We will discuss that the structure factor is not a good indicator of the jamming 
transition, as it predicts a similar structure immediately before and at $\phi_J$. We will 
introduce the local contact orientational order (LCOR), analogous to the local bond 
orientational order (LBOR), as a variable sensitive to jamming that quantifies the 
local structures of bidisperse packings.

This paper is organized as follows. In Sec.~\ref{SecII}, we briefly discuss the numerical simulation.
We define the concentration of small particles, $X_{\mathrm S}$, 
and discuss how the number of particles in each bidisperse mixture changes with $X_{\mathrm S}$. 
We also explain the simulation protocol used to determine the jammed structures. 
Sec.~\ref{SecIII} presents the first and second transitions for bidisperse packings. Here 
we present the method to obtain $\phi_J$. In Sec.~\ref{SecIV}, the structure factor 
is obtained to analyze the packing structure along the first and second transition lines. In 
Sec.~\ref{SecV}, we introduce the local contact orientational order, $\tilde{Q}_{\ell}$, to
quantify the local structure of the packings. Here, we investigate how $\tilde{Q}_{\ell}$ 
changes with $\phi$ and $X_{\mathrm S}$ for different $\delta$ values. We also show results 
of $\tilde{Q}_{\ell}$ for each configuration type to investigate their contribution to the 
jammed packing. Finally, we conclude with a summary and further discussion.

\section{Numerical simulation}
\label{SecII}

We perform 3D molecular dynamic simulations using MercuryDPM \citep{MercuryPage, weinhart2020fast} 
to study the role of small particles in the jammed structure of soft-sphere packings without gravity 
\cite{cundall1979discrete, petit2017contact, petit2018reduction}. The absence of gravity is essential 
for our observations. It allows small particles to have no contacts with the large particles 
and thus can undergo a collective transition upon high compression. In contrast, in the presence of gravity, 
small particles have already contacts with the jammed structure of large particles. These contacts make it
difficult to study the contribution of small particles on the jammed structure upon compression, and as a consequence,
any additional transition associated with small particles, given either by jamming density or LCOR, 
cannot be found. Newton's equation for each particle is solved numerically to predict its motion in time. 
$N = 6000$ particles are used to create a bidisperse packing, where a number of large, 
$N_{\mathrm L}$, and small, $N_{\mathrm S}$, particles with dimensionless radius 
$r_{\mathrm L}$ and $r_{\mathrm S}$ are considered. We choose the large particle radius as length 
scale, $x'_{u} = r'_{\mathrm{L}} = 1.5$, therefore, the dimensionless radius of large particles is 
$r_{\mathrm{L}} = r'_{\mathrm{L}}/x'_{u} = 1$, while for small particles, 
$r_{\mathrm{S}} = r'_{\mathrm{S}}/x'_{u} = r'_{\mathrm{S}}/r'_{\mathrm{L}}$, respectively. The prime
symbol represents the variable with units while the variable without prime is dimensionless.
These definitions above define the size ratio as $\delta = r'_{\mathrm{S}}/r'_{\mathrm{L}} = r_{\mathrm{S}} \in [0.15, 1]$. 
This means that any change in $\delta$ is due to a change in the small particle size. 
The mass scale is chosen as $m'_{u} = \rho'_{p} r_{\mathrm{L}}^{\prime 3}$, 
where $\rho'_{p} = 2000$ is the density of large and small particles and its 
dimensionless value is $\rho_{p} = 1$ since $\rho'_{u} = \rho'_{p}$. Therefore, the dimensionless mass 
of the large and small particles is $m_{\mathrm{L}} = \frac{4}{3}\pi$ and 
$m_{\mathrm{S}} = \frac{4}{3}\pi r_{\mathrm{S}}^{3} = m_{\mathrm{L}}\delta^{3}$.
The chosen time scale is $t'_{u} = (m'_{u}/\kappa'_{n})^{1/2}$ with $\kappa'_{n} = 10^{5}$ 
the normal stiffness. We choose here $\kappa_{n} = 1$, since $\kappa'_{u} = \kappa'_{n}$. Thus 
$t'_{u} = (\rho'_{p}/\kappa'_{n})^{1/2} r_{\mathrm{L}}^{\prime 3/2} \approx 0.26$.
The viscous damping used is $\gamma'_{n} = 1000$ and its dimensionless value is 
$\gamma_{n} = \gamma'_{n}/(\rho'_{u}\kappa'_{n} r_{\mathrm{L}}^{\prime 3}))^{1/2} \approx 0.038$.

\begin{figure}[t]
    
    \centering \includegraphics[scale=0.3]{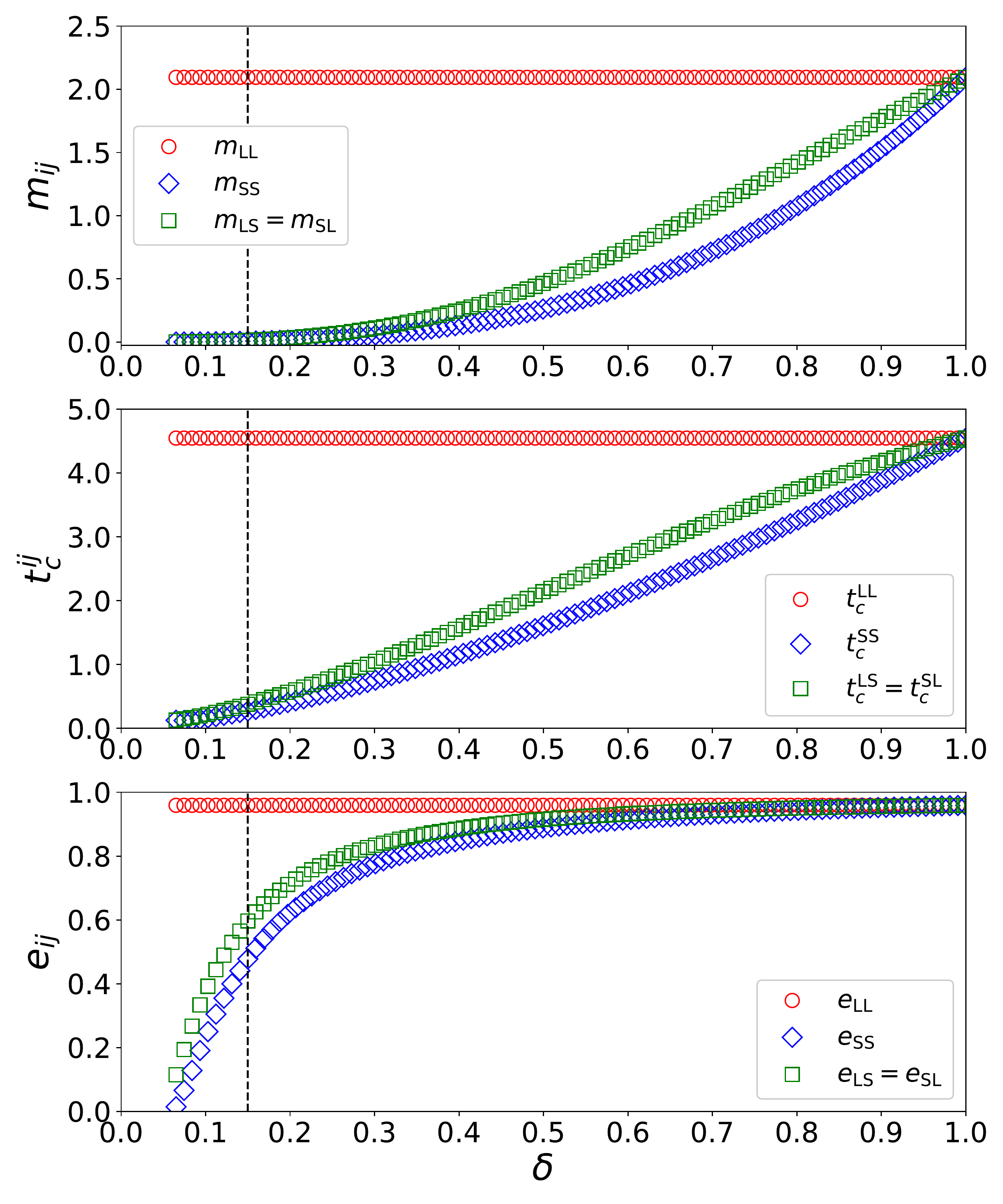}

    \caption{Variation of the dimensionless effective mass, $m_{\mathrm{ij}}$, the dimensionless
     contact time, $t_{c}^{ij}$, and the partial coefficient of restitution, $e_{ij}$, as a function of 
     the size ratio $\delta$. The dimensionless mass for each contact type is given by $m_{\mathrm{ LL }} = 2\pi/3$, 
     $m_{\mathrm{ SS }} = m_{\mathrm{ LL }} \delta^{3}$, and 
     $m_{\mathrm{ LS }} = m_{\mathrm{ SL }} = 2m_{\mathrm{ LL }}\delta^{3}/(1 + \delta^{3})$, 
     respectively. The dimensionless contact time is determined by 
     $t^{ij}_{c} =t^{\prime ij}_{c}/t^{\prime}_{u} = \pi/(\kappa_{n}/m_{ij} - (\gamma_n/m_{ij})^{2})^{1/2}$. 
     The dashed line represents the lowest value of $\delta = 0.15$ used in this work. Note that $e_{ij}$  
stops at $\delta \approx 0.06$ since imaginary values are obtained for $e_{\rm SS}$ and $e_{\rm LS}$ below it.}
	\label{parameters}
\end{figure}

The linear spring-dashpot model is used to model the contact between particles 
\cite{cundall1979discrete, petit2017contact, petit2018reduction, kumar2016tuning}.
For bidisperse packings, the effective mass, $m_{ij}$, the contact time, $t^{ij}_{c}$, 
and the coefficient of restitution, $e_{ij}$, depend on $\delta$, as can be seen in 
Fig.~\ref{parameters}. Since $e_{ij}$ depends on $m_{ij}$ and $t^{ij}_{c}$ via 
$e_{ij} = \exp \left ( - \gamma_{n}t_{c}^{ij}/2m_{ij} \right )$, the partial
coefficients of restitution for $\delta = 0.15$ are $e_{\mathrm{ LL }} = 0.95$, $e_{\mathrm{ SS }} = 0.48$, 
and $e_{\mathrm{ LS }} = e_{\mathrm{ SL }} = 0.60$, see the dashed line in Fig.~\ref{parameters}. 
However, as $\delta \to 1$, $e_{\mathrm{ SS }},e_{\mathrm{ LS }} \to e_{\mathrm{ LL }}$. 
This result indicates that the collisions of the SS and LS-SL configuration types are 
more elastic at high $\delta$. A background dissipation force is imposed on each particle velocity,
with constant dissipation $\gamma_{b} = \gamma_{n}$, to damp out the kinetic energy of the 
particles, especially at high $\delta$.

\begin{figure}[t]
	\centering \includegraphics[scale=0.3]{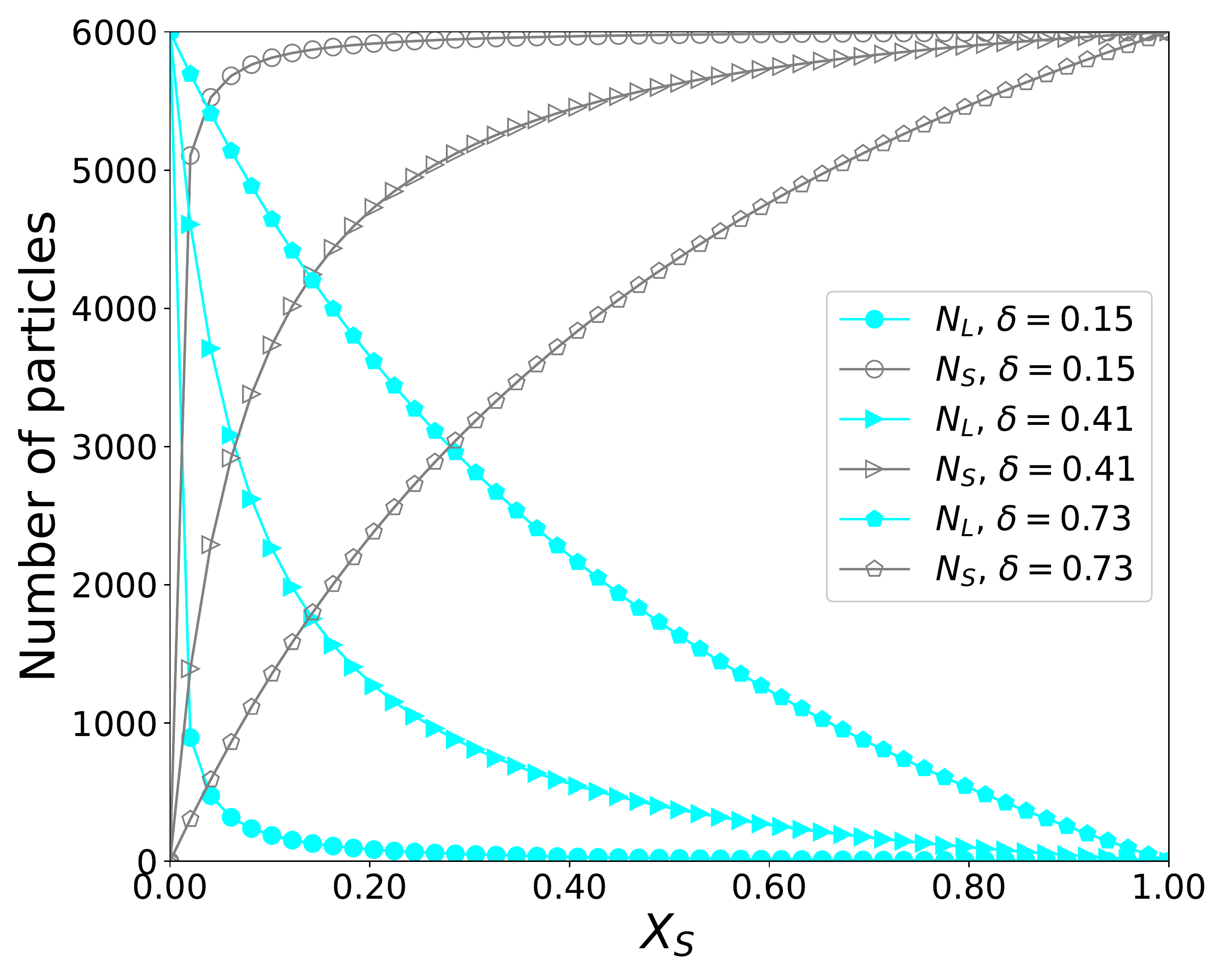}

	\caption{Number of large, $N_{\mathrm L}$, and small, $N_{\mathrm S}$, particles as a function of 
	$X_{\mathrm S}$ for three typical $\delta$. The total number of particles is fixed at $N = 6000$. 
	The intersection points represent the 50:50 mixtures at $X_{\mathrm S}(\delta = 0.73) \approx 0.28$, 
	$X_{\mathrm S}(\delta = 0.41) \approx 0.06$, 
	and $X_{\mathrm S}(\delta = 0.15) = 0.01$.}
	\label{protocol}
\end{figure}

\begin{figure}[t]

\centering \includegraphics[scale=0.32]{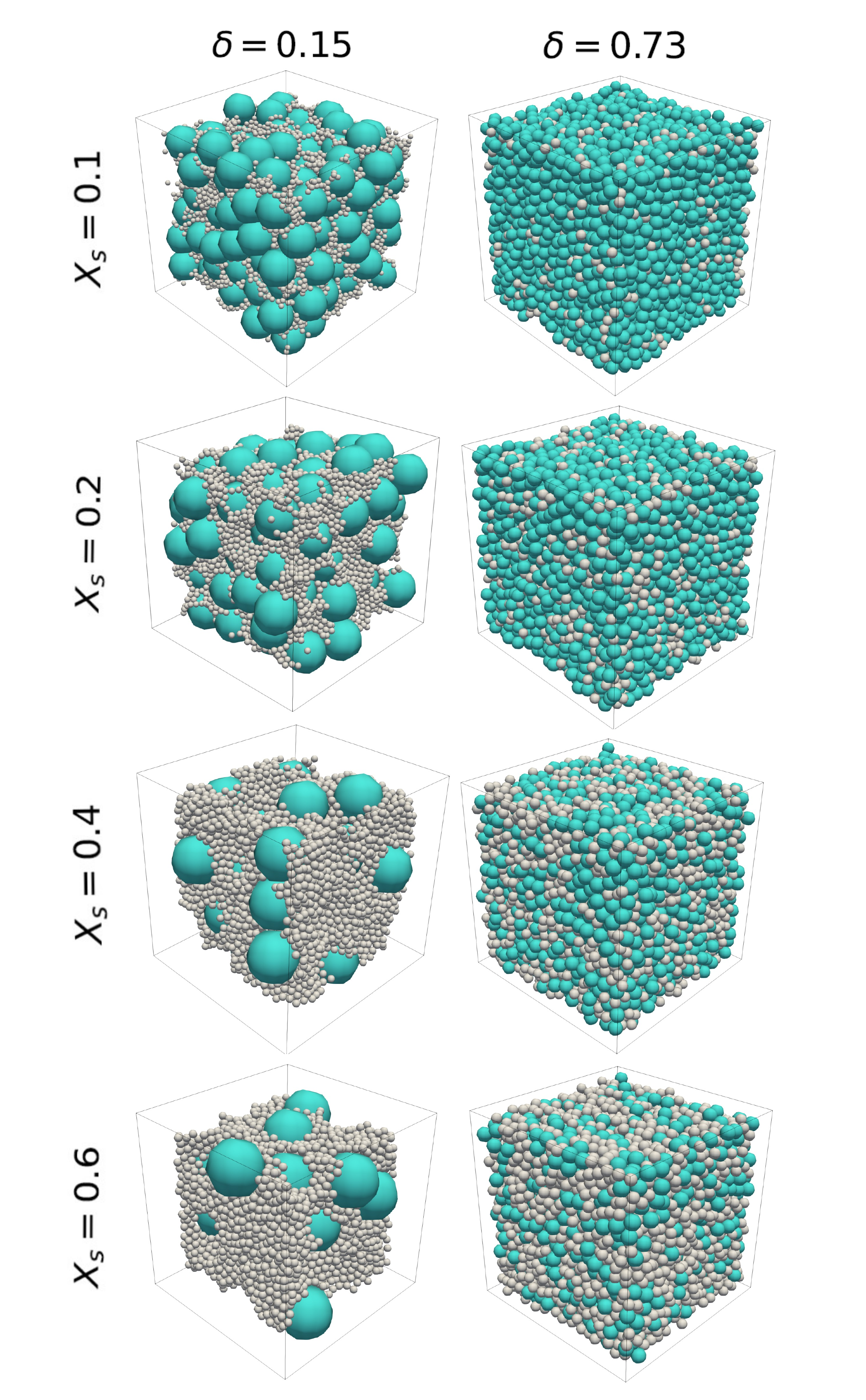}

\caption{Packing structures for two extremes $\delta$ at different $X_{\mathrm S}$. Cyan 
and white colors represent large and small particles, respectively. Each packing is shown at 
$\phi_{\mathrm{max}}$. Note that higher $\phi$ are obtained for structures with $\delta = 0.15$, 
they are shown enlarged for better illustration. This is the reason why large particles look bigger.}
\label{packings}
\end{figure}

A bidisperse packing formed by a given set of $N_{\mathrm L}$ and $N_{\mathrm S}$
is characterized by the size ratio $\delta$ and the volume concentration of the small particles, 
$X_{\mathrm S} = N_{\mathrm S} \delta^{3} / (N_{\mathrm L} + N_{\mathrm S} \delta^{3})$.
Fig.~\ref{protocol} shows the variation of $N_{\mathrm L}$ and $N_{\mathrm S}$ as a 
function of $X_{\mathrm S}$ for three typical values of $\delta$. 
For a fixed value of $\delta$ the number of small particles increases while the 
number of large particles decreases with $X_{\mathrm S}$. The intersection point, 
representing a packing with $N_{\mathrm L} = N_{\mathrm S} = N/2$, shifts to lower $X_{\mathrm S}$
values as $\delta$ decreases. This point corresponds to the 50:50 particle mixture studied previously in 
bidisperse systems using $\delta = 0.71$ \cite{ohern2003jamming, majmudar2007jamming}. 
Far below the intersection point ($X_{\mathrm S} \to 0$), the packing is formed by small 
particles in a sea of large particles. As $X_{\mathrm S}$ increases and approaches the intersection 
point, the numbers of small and large particles become of the same order of magnitude. Well above 
the intersection point ($X_{\mathrm S} \to 1$) few large particles are embedded in a sea 
of small ones. This can be seen in Fig.~\ref{packings} for typical bidisperse packing structures.

The initial configuration of any bidisperse packing is such that spherical particles of 
radius $r_{\mathrm L}$ and $r_{\mathrm S}$ are placed uniformly at random in a 3D box 
without gravity, allowing overlap between them, with an initial packing fraction of 
$\phi_{\mathrm{ini}} = 0.3$ and large uniform random velocities. Large overlaps 
lead to an initial peak in kinetic energy, but this is quickly damped by the background 
medium and collisions. Low density systems with high kinetic energy contribute to the rapid 
randomization of particles. The granular gas is then isotropically compressed to approach 
an initial direction-independent
configuration with the target packing fraction $\phi_0 < \phi_J$ that depends on 
$\delta$ and $X_{\mathrm S}$. Then a relaxation process of the system starts. Once such a process is complete, isotropic 
compression  (loading) begins, which ceases when $\phi = \phi_{\mathrm{max}}$. Then the 
isotropic decompression (unloading) process continues 10 times slower than the loading 
process until $\phi_{0}$ is reached again. In this way, the jamming density, $\phi_{J}$, 
along the decompression process is obtained.  
Other methods of strain control could be used \cite{donev2006binary, ohern2003jamming, chaudhuri2010jamming}, 
but they would not have any other effects since the deformation is performed quasi-statically.
After the simulation protocol is completed, the jamming density and the jammed structures of each 
bidisperse packing in the decompression branch are examined, since these values are less sensitive 
to the deformation rates \cite{goncu2010constitutive}. A detailed discussion of the contact model and
simulation procedure is given in Refs.~\cite{petit2020additional, petit2022bulk}.

\section{Jamming transition lines}
\label{SecIII}

\begin{figure}[t]
    
    \centering \includegraphics[scale=0.28]{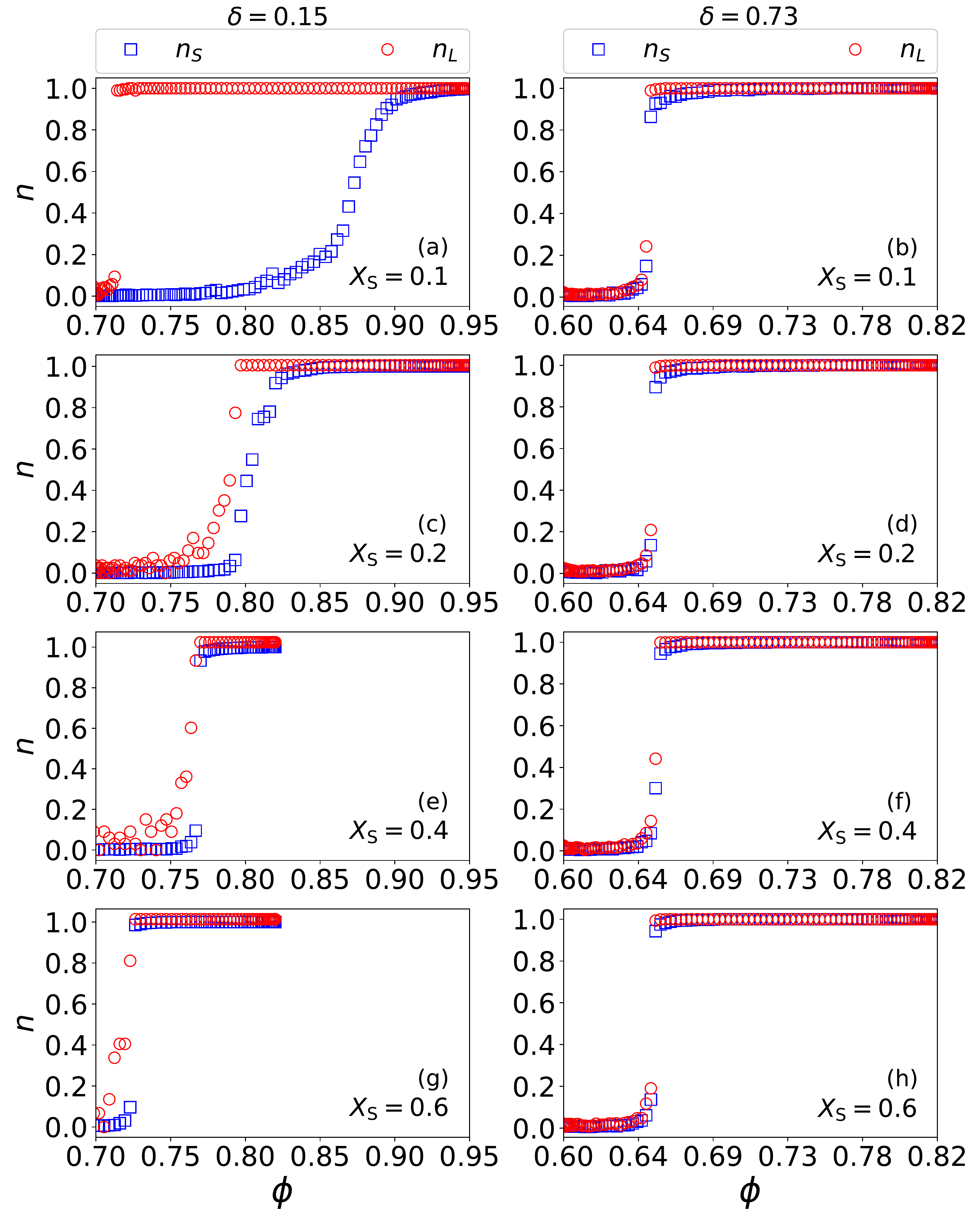}

    \caption{Fraction of large, $n_{\mathrm{L}}$, and small, $n_{\mathrm{S}}$, 
    particles in contact as a function of the packing fraction for different 
    combinations of $\delta$ and $X_{\mathrm S}$.}
	\label{fractpart_nu}
\end{figure}

In this section, we discuss how the jamming transition is achieved for each bidisperse packing.
We start by quantifying the fraction of large, $n_{\mathrm L} = N^{c}_{\mathrm{L}}/N$, 
and small particles, $n_{\mathrm S} = N^{c}_{\mathrm{S}}/N$, that contribute to the jammed 
structure as a function of $\phi$ at different $\delta$ and $X_{\mathrm S}$. $N^{c}_{\mathrm{L,S}}$ 
is the number of large and small particles in contact, while $N = N_{\mathrm{L}} + N_{\mathrm{S}}$
is the total number of particles in the system. 
Fig.~\ref{fractpart_nu} (right panel) shows a jump for $\delta = 0.73$, which represents a 
simultaneous contribution of both particle sizes to the jammed structure. Interestingly, 
$\phi_J$ is independent of $X_{\mathrm S}$ showing similar values to that of a monodisperse packing, 
$\phi_J^{\mathrm{mono}} \approx 0.64$. This finding is in line with the observation that a 
bidisperse packing with $\delta = 0.73$ can be used to break up a global crystallization obtained in 
monodisperse packings but otherwise behaves similar to it 
\cite{majmudar2007jamming, ohern2003jamming, morse2016geometric}. 
Although, as we will see in Sec.~\ref{SecV}, we still find a certain fraction of local 
ordered structures. In contrast, a significant 
decoupling between $n_{\mathrm{L}}$ and $n_{\mathrm{S}}$ is obtained at lower $X_{\mathrm{S}}$ for 
$\delta = 0.15$, see Fig.~\ref{fractpart_nu} (a,c). Such decoupling indicates that a large 
number of small particles are jammed at higher densities, which corresponds to similar 
behavior of large particles at low densities.

\begin{figure}[t]
    
    \centering \includegraphics[scale=0.28]{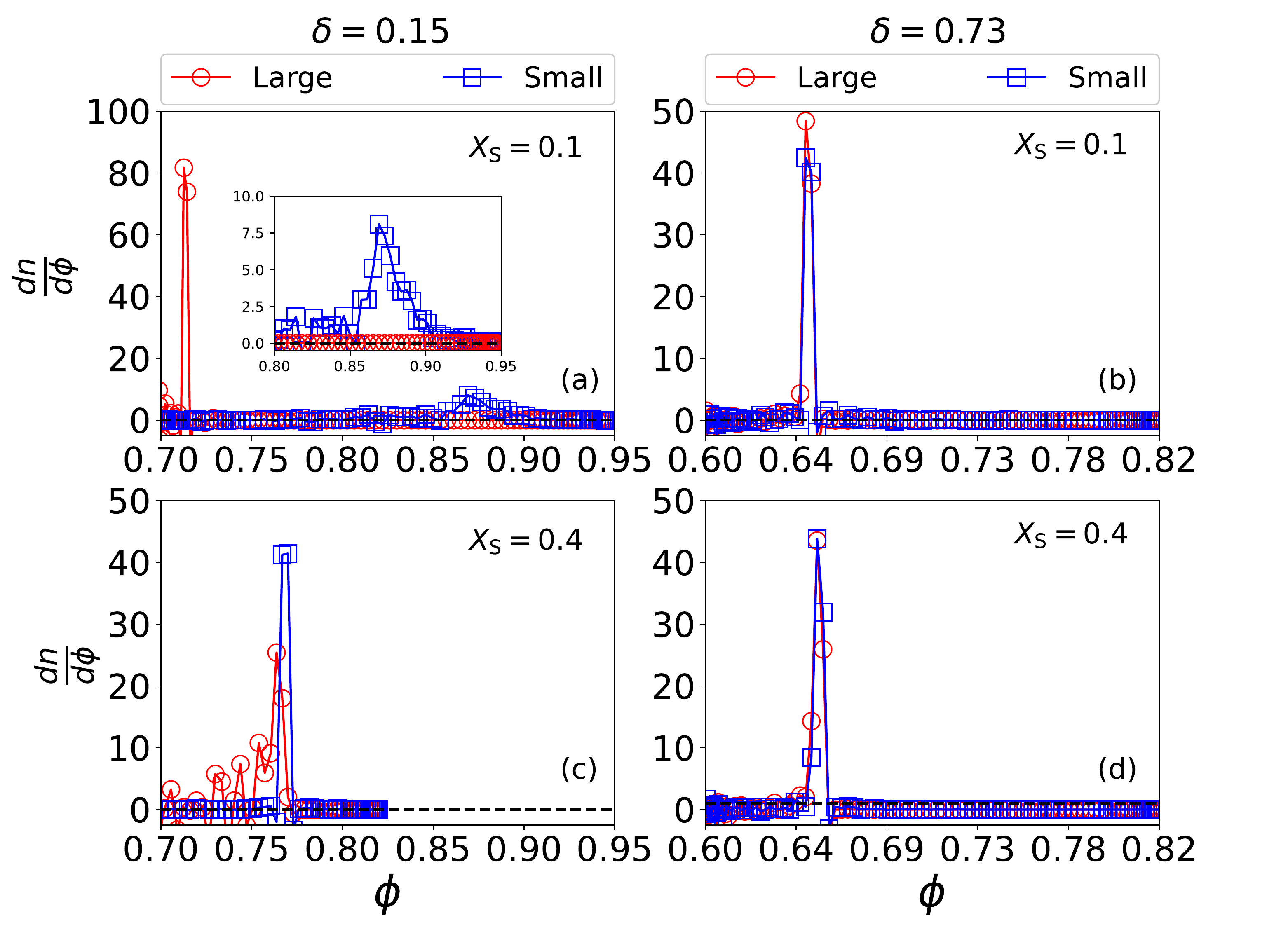}

    \caption{Derivative of $n_{\mathrm{L}}$ and $n_{\mathrm{S}}$ as 
    a function of the packing fraction for $\delta = 0.15$ and $\delta = 0.73$ at different 
    $X_{\mathrm{S}}$. The maximum value of each derivative is considered as the jamming density 
    of each particle size. The inset is a zoom-in of the maximum derivative of $n_{\mathrm{S}}$. }
	\label{Dev_frac_nu}
\end{figure}

To find the exact value of the jamming density at which $n_{\mathrm L}$ and $n_{\mathrm S}$ 
jump as a function of $\delta$ and $X_{\mathrm S}$, we calculate the derivative 
$\partial n_{\mathrm L}/ \partial \phi$ and $\partial n_{\mathrm S}/ \partial \phi$.
We used the five-point finite difference method with an accuracy of 
$\sim O(\Delta \phi^{4})$ to approximate the first derivative over the data shown in 
Fig.~\ref{fractpart_nu}. This method yields a value of $\phi_J$ for both fractions 
of large and small particles. Fig.~\ref{Dev_frac_nu} displays the derivative of $n_{\mathrm L}$ 
and $n_{\mathrm S}$ as a function of $\phi$, showing a characteristic peak (maximum derivative) 
at a value consistent with $\phi_J$. Note that for $\delta = 0.15$ and $X_{\mathrm S} = 0.1$ 
the peak for large particles is found at a much lower $\phi$, while a small peak 
is obtained at higher density for small particles, see Fig.~\ref{Dev_frac_nu} (a). The small peak 
in $n_{\mathrm S}$ is due to its smoother behavior compared to $n_{\mathrm L}$. Nevertheless, 
a critical density can be extracted representing the largest amount of small particles 
jammed, see the inset in Fig.~\ref{Dev_frac_nu} (a). This proves that the system undergoes 
a transition from a structure with predominantly large particles to one with the participation 
of both particle sizes. On the other hand, at higher $\delta$, it becomes clear that 
both particle sizes contribute simultaneously to the jammed structure, see Figs.~\ref{Dev_frac_nu} (b,\,d). 
Thus, using this method, 
one can extract the values of $\phi_J$ for the entire combination of $\delta$ and $X_{\mathrm S}$.

\begin{figure}[t]

    \centering \includegraphics[scale=0.34]{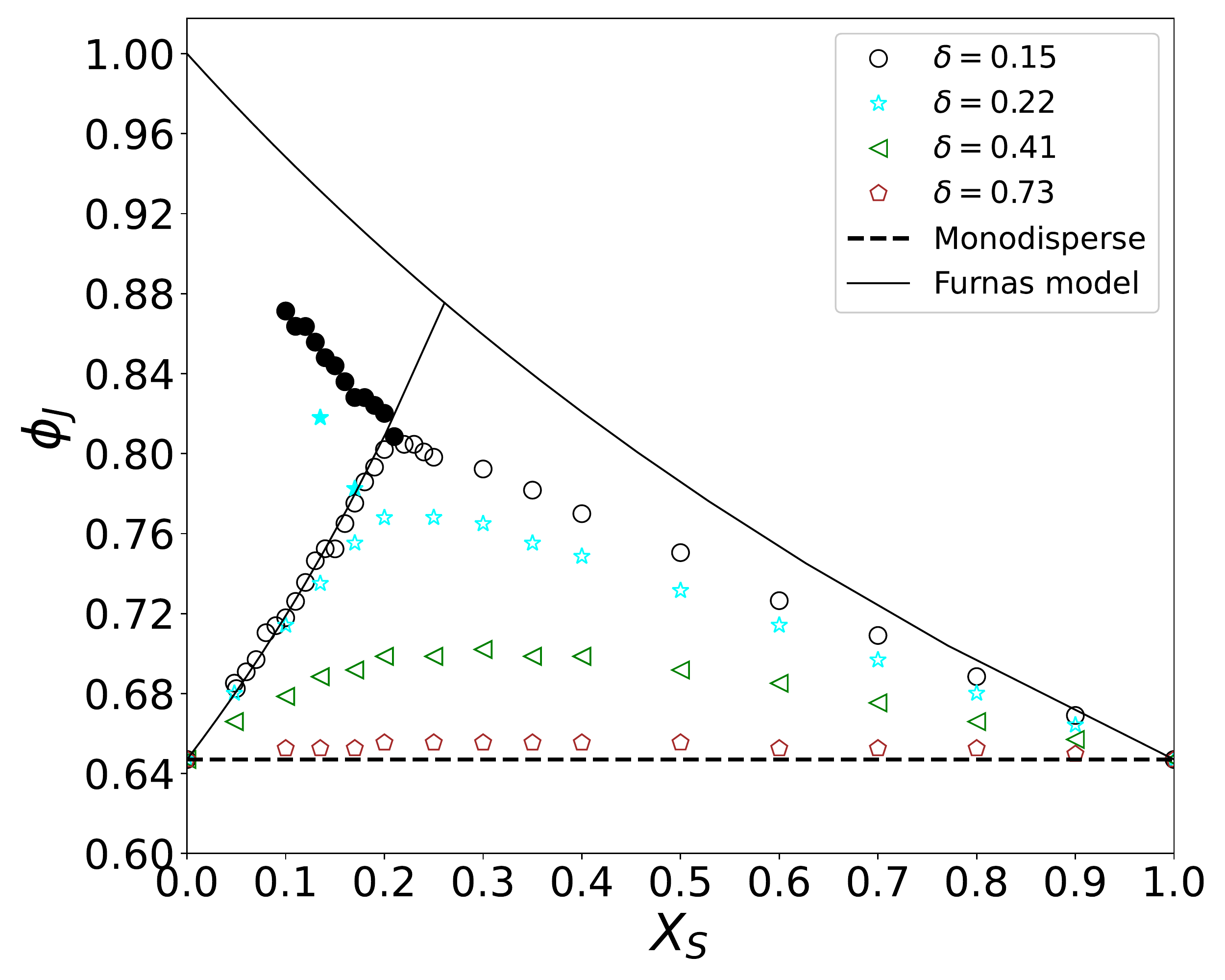}

 \caption{Jamming density, $\phi_{J}$, as a function of the concentration of small particles, $X_{\mathrm S}$, 
 for different values of the size ratio, $\delta$. The extreme $X_{\mathrm S}$ values (0 and 1) correspond to 
 monodisperse systems, which have a value of $\phi^{\mathrm{mono}}_{\mathrm{J}} \approx 0.64$, indicated by 
 the dashed horizontal line. The solid lines represent the Furnas model \cite{furnas1931grading}, 
 see the text for its explanation and ideas.  Open (solid) symbols represent the first (second) transition lines.}
  \label{jamming}
\end{figure}

Fig.~\ref{jamming} shows the $\phi_{J}$ values extracted by the method explained above as a 
function of $X_{\mathrm S}$ for some $\delta$ values. For $\delta = 0.15$, we find that the 
two lines meet at $X^{*}_{\mathrm S} \approx 0.21$ with $\phi_{J} \approx 0.80$. 
The superscript $\ast$ indicates the 
concentration of small particles where the second jamming transition line emerges, see solid circles in Fig.~\ref{jamming}. 
Such a point matches with the kink of the jamming lines, shifting to high $X_{\mathrm S}$ 
as $\delta$ increases, as also reported in Ref.~\cite{prasad2017subjamming}. For 
$X_{\mathrm S} < X^{*}_{\mathrm S}$, an increasing line of densities is observed for small $\delta$
as $X_{\mathrm S} \to 0$. Such a line is an extension of the transition where both large and small 
particles are jammed, having a particle mean overlap less than $1\%$ of the large particle radius, 
see Ref.~\cite{petit2022bulk}. The values of $\phi_J$ are compared with a model introduced by 
Furnas almost a century ago \cite{furnas1931grading} to predict the highest density of aggregates 
used in the production of mortar and concrete. This model states that $\phi_J$ can decouple at an 
extreme particle size ratio ($\delta \to 0$) into two limits that have a common point at $X^{*}_{\mathrm S}$. 
The lower limit considers an approximation where large particles dominate 
the jammed structure, while small particles are not considered because their number is not sufficient to play a role 
($0 \leq X_{\mathrm S} < X^{*}_{\mathrm S}$).  Thus, the jamming density is given by 
$\phi_J(X_{\mathrm S}) = \phi_J^{\mathrm{mono}}/(1 - X_{\mathrm S})$. 
The upper limit, both large and small 
particles participate in the jammed structure ($0 \leq X_{\mathrm S} \leq 1$). In this case, the 
number of small particles is large enough to drive some large particles into the jammed state. 
Therefore, $\phi_J$ is written by 
$\phi_J(X_{\mathrm S}) = \phi_J^{\mathrm{mono}}/(\phi_J^{\mathrm{mono}} + (1 - \phi_J^{\mathrm{mono}}) X_{\mathrm S})$.  
The Furnas model describes the trend of the data by following the values for low $X_{\mathrm S}$ 
corresponding to the first jamming state. It shows a maximum density of $\phi_J(X^{*}_{\mathrm S}) \approx 0.87$ 
at $X^{*}_{\mathrm S}=(1-\phi^{\mathrm{mono}}_{\mathrm{J}})/(2-\phi^{\mathrm{mono}}_{\mathrm{J}})\approx0.26$, 
which is in reasonable agreement with the value obtained here for $X^{*}_{\mathrm S} \approx 0.21$ 
at $\delta = 0.15$. The model also shows an additional transition line emerging where the two limits 
meet and end at a density of one. Such an additional line has not been considered in 
previous works when using the Furnas model, see 
Refs.~\cite{prasad2017subjamming,hopkins2011phase,biazzo2009theory,pillitteri2019jamming}.
The additional line resulting from our simulation data qualitatively follows 
the Furnas prediction and ends at $X_{\mathrm S}^{\circ} = 0.1$ for the lowest $\delta$, see Fig.~\ref{jamming}. 
The superscript $\circ$ marks the end-point of the extension line. The transition line ends at 
$X_{\mathrm S}^{\circ}$ since there is no jump in $n_{\mathrm{S}}$ 
for $X_{\mathrm S} < X_{\mathrm S}^{\circ}$, but rather this quantity increases continuously in this 
region and does not exhibit any features of a jump transition. This allows us to argue that the 
additional transition line terminates in an endpoint at a finite $X_{\mathrm S}^{\circ}$ that depends on $\delta$, 
see Ref.~\cite{petit2020additional}.

The jamming transition lines observed in Fig.~\ref{jamming} represent a more complete jamming diagram for 
bidisperse packings. Indeed, the second transition starts at a size ratio around $\delta = 0.22$ and becomes 
longer for smaller $\delta$. This particular value of $\delta$ coincides with the minimum size ratio, 
$\delta_{\mathrm{min}} \approx 0.225$, at which a small particle can fit into the gap left by large particles 
forming a tetrahedral structure, see Ref.~\cite{kumar2016tuning}. For $\delta > \delta_{\mathrm{min}}$, a small 
particle cannot fit into the gap left by the large particles in contact, destroying the tetrahedral structure 
and creating different local 
structures. For $\delta < \delta_{\mathrm{min}}$, the small particle is too small to
fit into the gap of the tetrahedral becoming a rattler in a system. Instead, a number of small 
particles are now needed to fill the gap in order to come in contact with large particles. In this case, 
other local structures are developed. This will be discussed in Sec.~\ref{SecV}.

So far, we have shown the dependence of $\phi_J$ on $\delta$ and $X_{\mathrm S}$. $\phi_J$ is 
enhanced for $\delta < 0.73$, especially at very low $\delta$ and low $X_{\mathrm S}$, showing 
a second transition. The differences in jamming densities between the extreme size 
ratios are remarkable. For example, one would have expected a substantial variation of $\phi_J$ 
for $\delta = 0.73$, since this is far from the monodisperse case of only large, $\delta = 0$, 
and only small particles, $\delta = 1$. Instead, $\phi_J$ appears to be constant for $\delta = 0.73$ 
and hardly varies above $\phi^{\mathrm{mono}}_{\mathrm{J}}$ independent of $X_{\mathrm S}$. 
This means that packings with $\delta > 0.73$ would have $\phi_{J}$ values close to $\phi^{\mathrm{mono}}_{\mathrm{J}}$ 
and probably similar properties as a monodisperse packing. Examining the jammed structure of the packings as a 
function of $\delta$ and $X_{\mathrm S}$ may provide better insight into the values 
of $\phi_J$ for $\delta = 0.73$. In addition, it is important to understand how the 
jammed structure evolves as the system approaches jamming and how it changes along the first 
and second jamming transitions. The following sections are devoted to the study of the structure of 
jammed bidisperse packings along the jamming transition lines.

\section{Structure factor analysis}
\label{SecIV}

To understand how the structure of a bidisperse packing changes with $X_{\mathrm{S}}$ and $\delta$, 
we calculate the total and partial structure factors, $S(q)$, at $\phi_J$. This allows 
exploring the structural contribution that each configuration type has in the jammed 
packing. A general definition of the partial $S(q)$ is

\begin{equation}
\begin{split}
S_{\mathrm{\nu\beta}}(q)\ &= \frac{1}{N} \sum_{i=1}^{N_{\mathrm{\nu}}} \sum_{j=1}^{N_{\mathrm{\beta}}}
							   \mathrm{cos}(\mathbf{q}\cdot \mathbf{r}^{i}_{\mathrm{\nu}}) \mathrm{cos}(\mathbf{q}\cdot \mathbf{r}^{j}_{\mathrm{\beta}})\\  	
							 & + \frac{1}{N} \sum_{i=1}^{N_{\mathrm{\nu}}} \sum_{j=1}^{N_{\mathrm{\beta}}} 
							   \mathrm{sin}(\mathbf{q}\cdot \mathbf{r}^{i}_{\mathrm{\nu}}) \mathrm{sin}(\mathbf{q}\cdot \mathbf{r}^{j}_{\mathrm{\beta}})  
\end{split}
\label{ecu2}
\end{equation}

\noindent where $\nu,\,\beta \in \{\mathrm{L},\mathrm{S}\}$ and the sum runs over all $\nu$ 
and $\beta$ particles.
Therefore, the total $S(q)$ can be decomposed in terms of configuration types: 
$S_{\mathrm{ LL }}(q)$, $S_{\mathrm{ SS }}(q)$, $S_{\mathrm{ LS }}(q)$, and $S_{\mathrm{ SL }}(q)$
as shown in Ref.~\cite{xu2010effects}. By symmetry, we obtain that
$S_{\mathrm{ LS }}(q) = S_{\mathrm{ SL }}(q)$. Thus, we show the structure factor of only 
one term and call it $S_{\mathrm{mix}}(q)$. The term ``mix" 
is only used in this section to highlight the equivalence of $S(q)$ for SL and LS. As we will 
explain in Sec.~\ref{SecV}, SL and LS contact configurations are differently treated when calculating 
LCOR, thus the term ``mix" is no longer used. Therefore, the total structure factor is then written as 
$S(q) = S_{\mathrm{ LL }}(q) + S_{\mathrm{ SS }}(q) + 2S_{\mathrm{mix}}(q)$.

\begin{figure}[t]
\centering \includegraphics[scale=0.24]{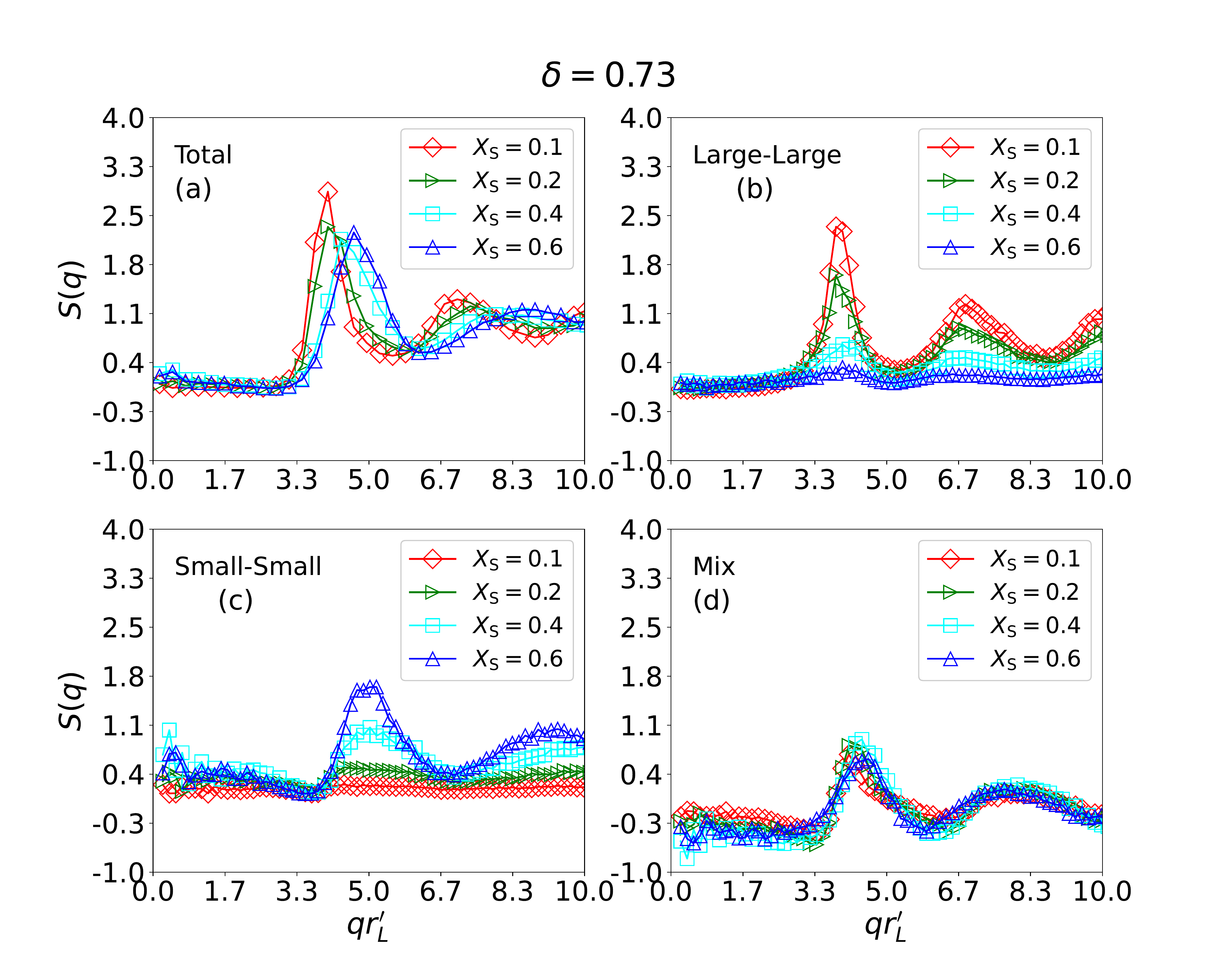}
 \caption{$S(q)$ vs $qr'_{\mathrm{L}}$ at $\phi_J$ for $\delta = 0.73$ at different $X_{\mathrm{S}}$. $S(q)$ of 
 (a) Total, (b) Large-Large, (c) Small-Small, and (d) Mixed particle configurations.}
  \label{plot_Sk_SRp7321}
\end{figure}

Fig.~\ref{plot_Sk_SRp7321} shows the total and partial structure factors at jamming
for $\delta = 0.73$ at different $X_{\mathrm{S}}$. We obtain that LL dominates over SS and 
mix configurations for the lowest $X_{\mathrm{S}}$, see Figs.~\ref{plot_Sk_SRp7321} (b)-(d). 
As $X_{\mathrm{S}}$ increases, SS begins to dominate the structure over the other configu\-ration 
types. This is evident as the number of small particles increases with $X_{\mathrm{S}}$. 
The exchange of the configuration type in the dominance of the packing structure marks a 
structural change above a certain $X_{\mathrm{S}}$. We think that this occurs at 
$X_{\mathrm{S}}(\delta = 0.73) \sim 0.28$ since it corres\-ponds to the 50:50 particle mixture 
of the packing, see Fig.~\ref{protocol}. On the other hand, $S_{\mathrm{mix}}(q)$ appears 
to be independent of $X_{\mathrm{S}}$, suggesting that it has no effect on the overall structure 
factor. The meaning of the negative value in $S_{\mathrm{mix}}(q)$ 
indicates anticorrelated density fluctuations between small and large particles at 
long distances ($q \to 0$), i.e., high density fluctuations of small particles correspond 
to low density fluctuation of large ones. For positive values,  
the density fluctuations of large and small particles need to be mostly in sync and so 
are correlated.

The total $S(q)$ shows a gradual change due to the structural transition that the 
LL and SS configurations undergo, causing the system to explore different local structures 
as $X_{\mathrm{S}}$ varies. Despite $S_{\mathrm{ LL }}(q)$ and $S_{\mathrm{ SS }}(q)$ show a 
significant change with $X_{\mathrm{S}}$, the whole $S(q)$ shows similar structure factors, i.e., 
similar jammed structures are obtained where their peaks become wider and shifted for high $q$ 
as $X_{\mathrm{S}} \to 1$. We think that the similar structures 
might be responsible for the similar jamming densities observed in Fig.~\ref{jamming} for 
$\delta = 0.73$. This indicates that the packing structure influences the jamming density, as 
it was recently shown in Ref.~\cite{blumenfeld2021disorder}, where 
the random close packing in monodisperse packings can be theoretically calculated by 
considering only specific local disorder arrangements of particles.

\begin{figure}[t]
\centering \includegraphics[scale=0.24]{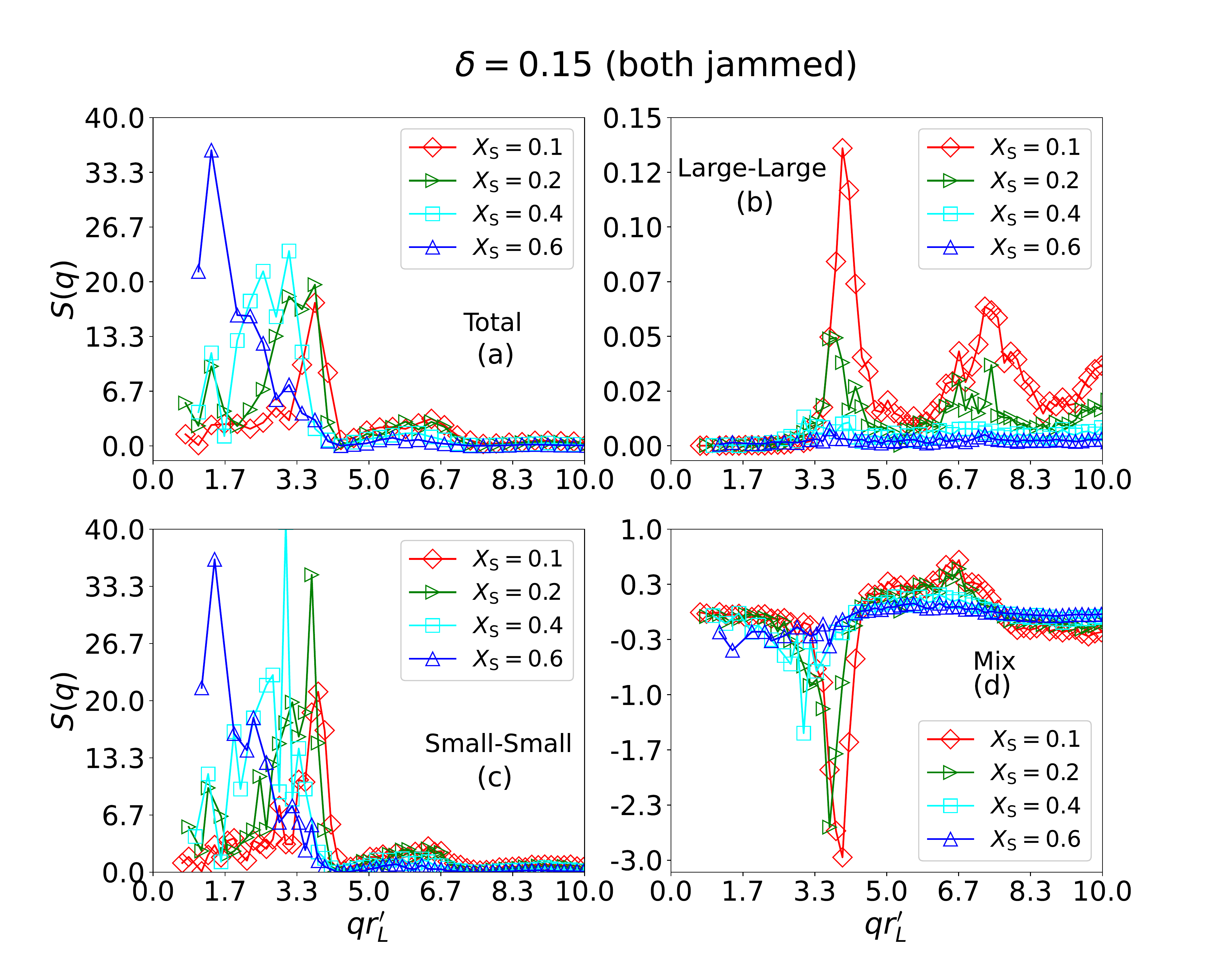}
 \caption{$S(q)$ vs $qr'_{\mathrm{L}}$ for $\delta = 0.15$ at different $X_{\mathrm{S}}$ along
 the transition line where both species are jammed. $S(q)$ of 
 (a) Total, (b) Large-Large, (c) Small-Small and (d) Mixed particle configurations. }
  \label{plot_Sk_both_SRp1548}
\end{figure}

For $\delta = 0.15$, the structure factor is different from $\delta = 0.73$. 
At the first transition, the total $S(q)$ is given by $S_{\mathrm{LL}}(q)$ since the jammed
structure consists only of large particles (data not shown). On the other hand, 
at the second transition, both large and small particles are jammed. The total $S(q)$
is dominated by SS over LL and mix configurations types, see the 
magnitude of the y-axis of Figs.~\ref{plot_Sk_both_SRp1548} (b)-(d). This dominance rises 
as $X_{\mathrm{S}}$ increases, leading to a packing structure formed mostly of small 
particles with some contribution from mix configurations, see Fig.~\ref{plot_Sk_both_SRp1548} (d).
The peaks shown by $S_{\mathrm{SS}}(q)$ at long wavelengths ($q \lesssim 3.3$), 
see Fig.~\ref{plot_Sk_both_SRp1548} (c), can be due to (i) the formation of ordered regions 
between small particles, which is possible because there are many more small particles than
large ones and (ii) high density fluctuations of small particles compared to large ones. 
Few large particles are surrounded by many small ones at large scales. This is also observed 
in Fig.~\ref{plot_Sk_both_SRp1548} (d), where a huge disparity between large and small 
particle number at a large distance gives rise to negative 
values in $S_{\mathrm{LS}}(q)$, leading to anticorrelated density fluctuations.

\begin{figure}[t]
\centering \includegraphics[scale=0.205]{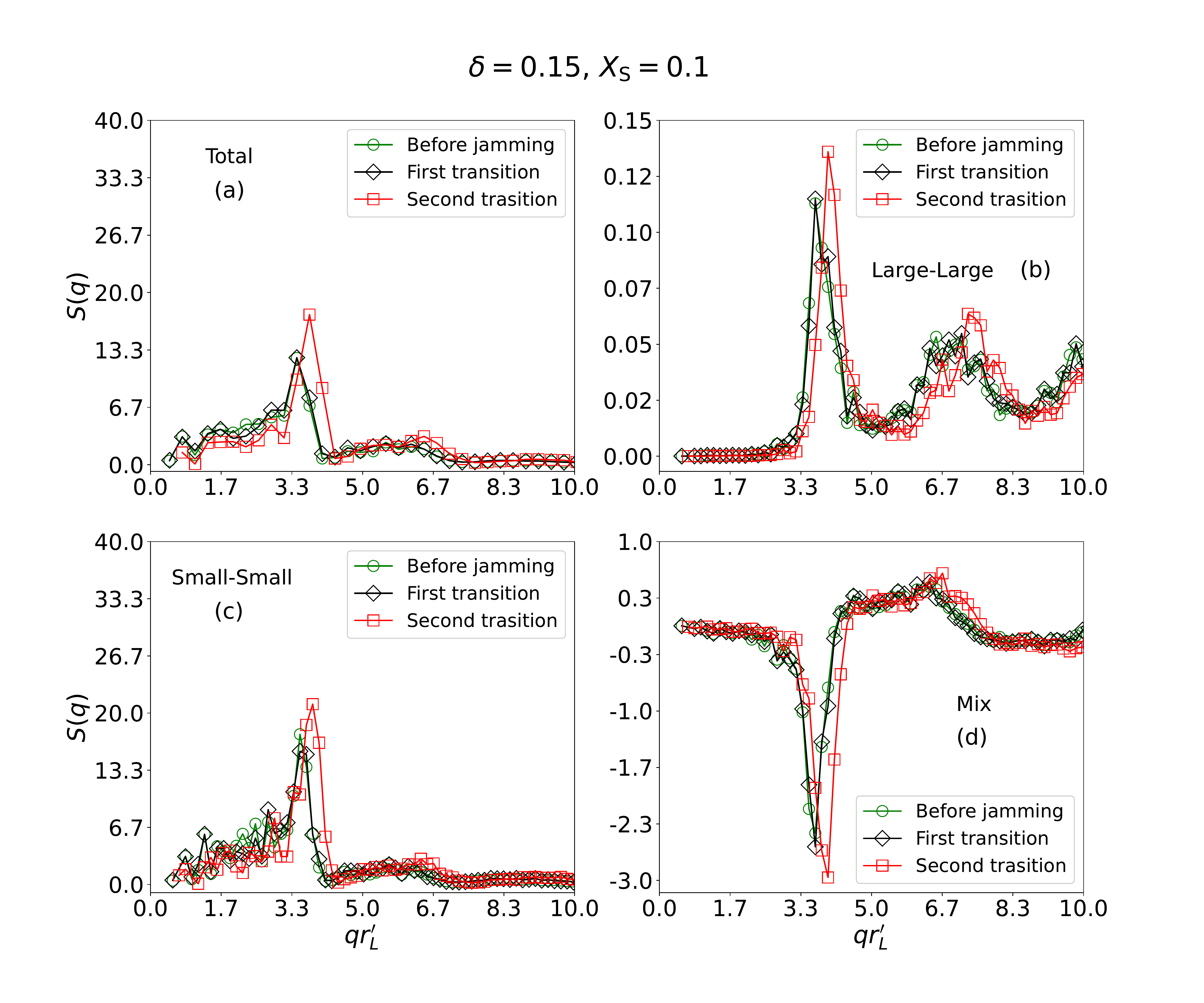}
 \caption{Comparison of $S(q)$ before ($\phi = 0.7143$), at the first 
 ($\phi_J = 0.7160$) and second jamming transition ($\phi_J = 0.8712$) for $\delta = 0.15$ 
 at $X_{\mathrm{S}} = 0.1$. $S(q)$ of (a) Total, (b) Large-Large, (c) Small-Small, 
 and (d) Mixed particle configurations.}
  \label{plot_Sk_compa_SRp1548}
\end{figure}

Fig.~\ref{plot_Sk_compa_SRp1548} shows the variation of total and partial $S(q)$
before, at the first and at the second jamming transition for $\delta = 0.15$ at 
$X_{\mathrm{S}} = 0.1$. The $S(q)$ shown at the first and second transitions differ 
as a consequence of further compression, which causes small particles to jam with the 
jammed structure of large particles and to form high 
crystallized regions of small particles, see the peaks in 
Fig.~\ref{plot_Sk_compa_SRp1548} (a,c) and particle configurations in Fig.~\ref{packings}. 
However, it is difficult to interpret such a 
difference as an indication of a jammed transition. On the other hand, the $S(q)$ 
immediately before and at the first jamming transition are identical to each other, 
which could be interpreted as the same packing structure. Note that the difference 
in the packing fraction between the structure before and the structure at 
the first jamming transition is $\Delta \phi \approx 10^{-3}$. This small value
does not make much difference between the total and partial $S(q)$ when going from 
a loose to the first jamming state. The reason for this is that all particles in the system 
are used to calculate the structure factor, regardless of whether the 
particles are in contact or not. This makes it difficult to distinguish whether 
the structure is present before, at the first, or even 
at the second jamming transition. In the next section, we will 
introduce and examine a sensitive variable that can distinguish the structural features of 
a jamming transition.

\section{Local contact orientational order $\tilde{Q}_{\ell}$}
\label{SecV}

In the previous section, we showed that $S(q)$ before jamming is identical to $S(q)$ 
at the first transition, suggesting that there is no difference 
in structure between them. As $\phi \to \phi_J$, a loose granular packing with a 
non-contacting structure approaches the jamming state. At $\phi = \phi_J$, the 
packing undergoes a structural transition defined as a jammed structure, which is not 
accounted for by the structure factor. To understand the evolution of a jammed structure, 
and even more to distinguish the structures along the first and second transitions given in 
Fig.~\ref{jamming}, a variable sensitive to each structural feature at 
jamming is needed. This section is dedicated to the introduction of a variable that is not only 
able to predict structural changes in the packing, but also to quantify the contribution 
of specific local structures formed by each configuration type in the jammed structure.

\subsection{Definition of $\tilde{Q}_{\ell}$}

A variable that has been used to study the structure and measure crystallinity in supercooled 
liquids and metallic glasses is the bond orientational order (BOR), $Q_{\ell}$, see 
Ref.~\cite{steinhardt1983bond}. It is determined by summing the spherical harmonics of 
degree $\ell$ of all bonds in the system. Here, a bond is defined by the connection of 
each particle center $i$ with the center of its nearest neighbors $j$. A local measure 
of this variable, $Q_{\ell,\rm{local}}$, was proposed in Refs.~\cite{steinhardt1983bond, 
kansal2002diversity} as a more accurate measure for identifying local structures. It 
determines the BOR on each particle and then is averaged over all particles. 
However, $Q_{\ell,\rm{local}}$ depends on the method used for nearest neighbor 
detection. For instance, the radial distribution function with a cutoff $r_{\rm{max}}$ is
used \citep{steinhardt1983bond, martin2008influence, duff2007shear}, however, it leads to different $Q_{\ell,\rm{local}}$ values when 
$r_{\rm{max}}$ is varied. Delauney triangulation method
\cite{fortune1995voronoi, kansal2002diversity}, morphometric neighbourhood \cite{mickel2013shortcomings}
and an extension of the morphometric neighborhood applied to noisy structures 
\cite{haeberle2019distinguishing} has also been applied for identifying nearest neighbors. 
Due to neighborhood ambiguity, $Q_{\ell,\rm{local}}$ is not uniquely defined. Here, we introduce an alternative definition of the bond orientational 
order. Instead of using a special detection method to find the nearest neighbors, we use the 
contacts between particles to define the \emph{local contact orientational order} (LCOR), 
$\tilde{Q}_{\ell}$. In this way, the neighbors of a particle $i$ are already defined by their 
contacts. The definition of $\tilde{Q}_{\ell}$ states that before jamming when no jammed 
structure has yet formed, zero values of $\tilde{Q}_{\ell}$ must be obtained. However, 
there is a possibility that an isolated accumulation of contact particles will yield a 
nonzero but low value of $\tilde{Q}_{\ell}$. Therefore, LCOR abandons the definition of a recent work 
\cite{mizuno2020structural}, in which the local bond orientational order at jamming is 
zero for highly amorphous packings. In our work, amorphous packings 
are characterized by the onset of the jamming structure where $\tilde{Q}_{\ell}$ is not 
necessarily zero but has a finite low value. While high values of $\tilde{Q}_{\ell}$ represent 
an ordered packing.
For dense packings where $\phi \gg \phi_J$, we expect BOR to be equal to LCOR, 
$Q_{\ell,\rm{local}} = \tilde{Q}_{\ell}$, since the detection method used in BOR can 
already identify those $j$ particles in contact with $i$ particle as nearest neighbors.

The local contact orientational order is then calculated by

\begin{equation}
\tilde{Q}_{\ell} = \frac{1}{N}\sum_{i = 1}^{N} 
				   \bigg( \frac{4\pi}{2\ell + 1} 
				   \sum_{m = -\ell}^{\ell} \bigg| \frac{1}{N_{c}^{i}} 
				   \sum_{j = 1}^{N_{c}^{i}} Y_{\ell m}(\theta_{j}, \varphi_{j})\bigg|^{2}
				   \bigg)^{1/2},
\label{ecu3}
\end{equation} 

\noindent where $Y_{\ell m}(\theta_{j}, \varphi_{j})$ is the spherical harmonics of 
degree $\ell$ and of order $m$, $\theta_{j}$ and $\varphi_{j}$ 
are the polar and azimuthal angles formed by $i$ particle center between their $j$ contacts with 
respect to the $z$ and $x$ axes, respectively. $N_{c}^{i}$ is the number of contacts of 
$i$ particle and $N$ is the total number of particles.

\subsection{Frequency distribution of $\tilde{Q}_{6}$}

To get a first insight into the structures of the jammed bidisperse granular packings, 
we determine $\tilde{Q}_{6}$ for each particle $i$ in the system. In this way, we 
can distinguish the LCOR of large particles from that of small particles. The reason for 
calculating $\tilde{Q}_{6}$ is because it can be used to quantify possible six-fold 
local crystal structures that form between particles of the same size. For example, it has 
been shown that the packing structure of particles of one size tends to form local HCP and 
FCC structures upon compression, becoming more frequent for denser packings \cite{clarke1993structural, klumov2011structural, 
klumov2014structural, hanifpour2015structural}. In particular, the distribution of the bond 
orientational order shows characteristic peaks at $\tilde{Q}_{6}^{\mathrm{HCP}} = Q^{\mathrm{HCP}}_{6,\mathrm{local}} = 0.48$ 
and $\tilde{Q}_{6}^{\mathrm{FCC}} = Q^{\mathrm{FCC}}_{6,\mathrm{local}} = 0.57$, consistent with the dominance of local 
HCP and FCC structures \cite{klumov2011structural, klumov2014structural, hanifpour2015structural}. 
Here, we assume that LCOR must exactly match with BOR for HCP and FCC structures.  
With this background, we can study how large and small particles are packed according to a 
six-fold lattice as a function of $\delta$ and $X_{\mathrm{S}}$.

\begin{figure}[t]

\centering \includegraphics[scale=0.42]{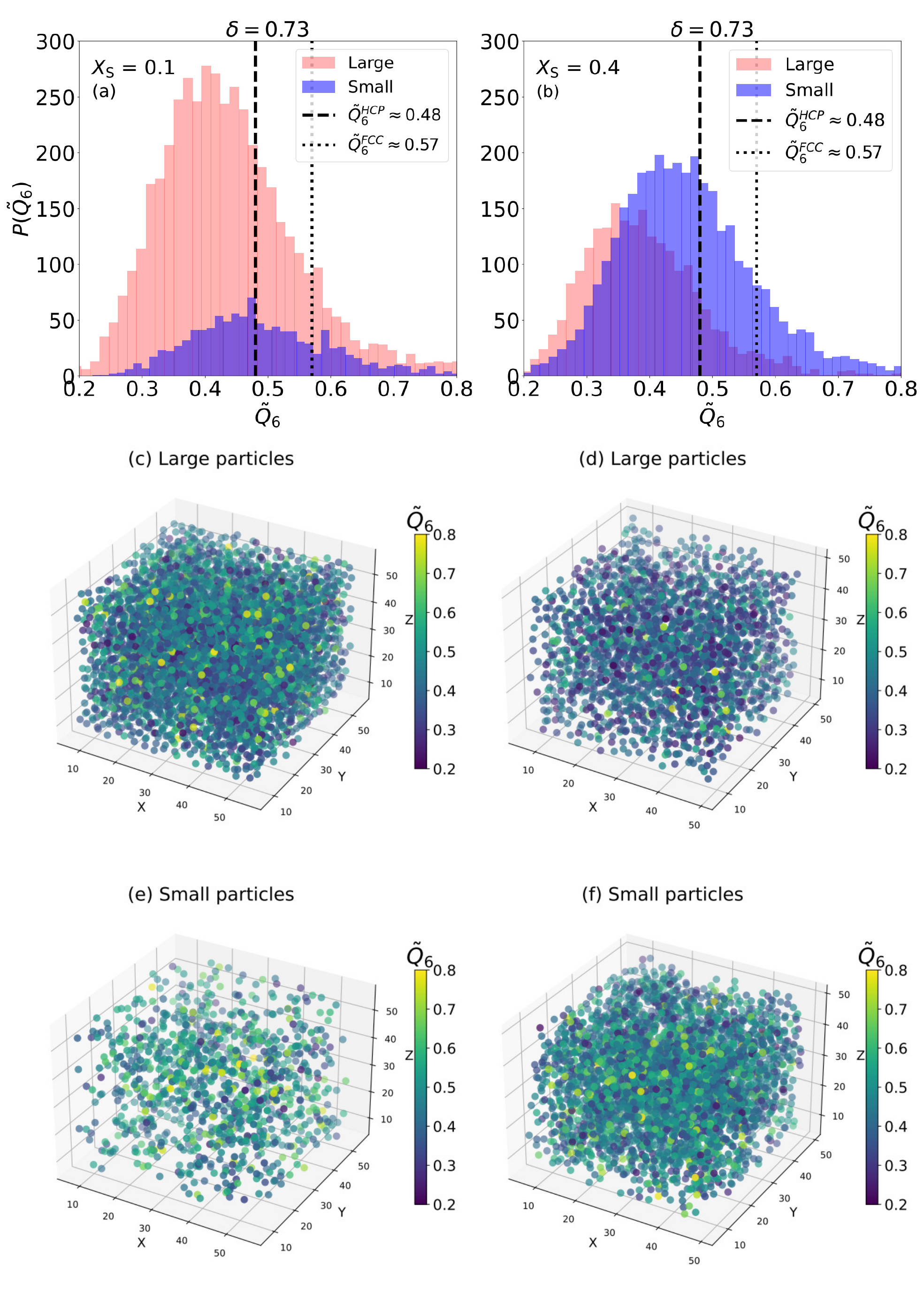}

 \caption{Frequency distribution, $P(\tilde{Q}_{6})$, of large and small particles at $\phi_J$ for 
 $\delta = 0.73$ at (a) $X_{\mathrm{S}} = 0.1$ and (b) $X_{\mathrm{S}} = 0.4$. The dashed and dotted lines
 represent $\tilde{Q}^{\mathrm{HCP}}_{6} = 0.48$ and $\tilde{Q}^{\mathrm{FCC}}_{6} = 0.57$, respectively. 
 (c)-(f) Configurations of large and small particles at $\phi_J$. Each dot 
 represents the center of a particle and the color indicates the magnitude of $\tilde{Q}_{6}$. The lowest value of $\tilde{Q}_{6}$ 
 (dark color) represents a disordered lattice, while the highest value (light color) is an 
 ordered one. The dots are barely transparent to reveal the structure behind them.}
  \label{HeatmapPlots_SRp7321}
\end{figure}

\begin{figure}[t]

\centering \includegraphics[scale=0.42]{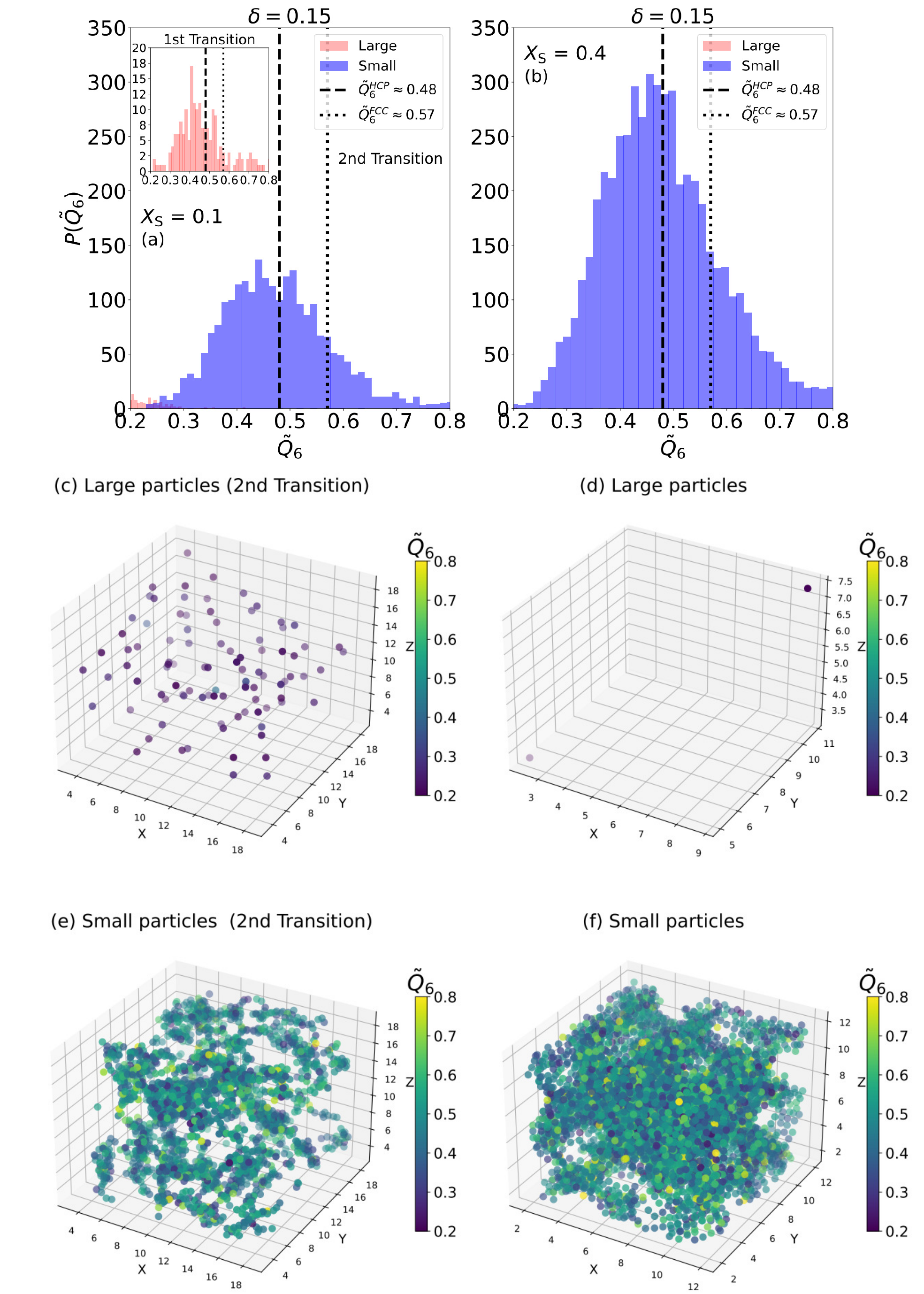}

 \caption{Frequency distribution, $P(\tilde{Q}_{6})$, of large and small particles at $\phi_J$ for 
 $\delta = 0.15$ at (a) $X_{\mathrm{S}} = 0.1$ at the second transition and (b) $X_{\mathrm{S}} = 0.4$. 
 The dashed and dotted lines represent $\tilde{Q}^{\mathrm{HCP}}_{6} = 0.48$ and $\tilde{Q}^{\mathrm{FCC}}_{6} = 0.57$, 
 respectively. The inset in (a) represents $P(\tilde{Q}_{6})$ at the first jamming transition.
(c)-(f) Configurations of large and small particles at $\phi_J$. Each dot represents the center of 
 a particle and the color indicates the magnitude of $\tilde{Q}_{6}$. The lowest value of $\tilde{Q}_{6}$ 
 (dark color) represents a disordered lattice, while the highest value (light color) is an 
 ordered one. The dots are barely transparent to reveal the structure behind them. }
  \label{HeatmapPlots_SRp1548}
\end{figure}

The values of $\tilde{Q}_{6}$ for large and small particles are used to construct 
independent frequency distributions, $P(\tilde{Q}_{6})$. Such distribution is shown in 
Fig.~\ref{HeatmapPlots_SRp7321} (a,\,b)  at $\phi_J$ for $\delta = 0.73$ at two relevant $X_{\mathrm{S}}$ 
values. $P(\tilde{Q}_{6})$ expresses the population 
of large and small particles with $\tilde{Q}_{6}$ within the jammed structure. In general, the distributions follow 
a Gaussian-like behavior, with their mean value depending on the particle size. For $X_{\mathrm{S}} = 0.4$ 
the structure is dominated by small particles, while for $X_{\mathrm{S}} = 0.1$ the large particles predominate. 
In both cases, $P(\tilde{Q}_{6})$ exhibits a fraction of local HCP and FCC structures formed by both large 
and small particles that change with $X_{\mathrm{S}}$, see the dashed and dotted lines in 
Figs.~\ref{HeatmapPlots_SRp7321} (a,\,b). For $\delta = 0.15$ a similar explanation can be given, but in this case, 
small particles always dominate the jammed structure over large ones, see Figs.~\ref{HeatmapPlots_SRp1548} (a,\,b). 
At $X_{\mathrm{S}} = 0.1$, the system undergoes several jamming transitions during compression. The first jamming 
transition is caused by large particles at $\phi_J \approx 0.71$, where $P(\tilde{Q}_{6})$ is 
indicated in the inset of Fig.~\ref{HeatmapPlots_SRp1548} (a). This shows that the packing is not fully ordered 
despite the formation of some local HCP and FCC structures. Instead, it shows a wide range of local structures. 
At the second transition, $\phi_J \approx 0.87$, the small particles dominate the jammed structure since
they disrupt the jammed structure of large ones resulting in less contacts of LL, giving rise to  
$0.2 \leq \tilde{Q}_{6} \leq 0.3$. As $X_{\mathrm{S}}$ increases, large 
particles are less present in the system, leading to a monodisperse packing of small particles. 
Fig.~\ref{HeatmapPlots_SRp1548} (b) shows this scenario, where the mean of the distribution coincides with 
$\tilde{Q}^{\mathrm{HCP}}_{6}$, indicating that most small particles form hexagonal local structures.
We also find that the population of local HCP and FCC structures increases with $X_{\mathrm{S}}$, suggesting 
that the local order of the packing increases with the concentration of small particles.

For a deeper understanding of the local structure, the configuration of large and small 
particles at jamming are separately depicted using dots with specific colors. Each dot is placed at 
each particle center while its color represents the value of $\tilde{Q}_{6}$ of the particle.
Dark colors represent the local disordered surrounding of the particles, while
light colors correspond to the local ordered ones. Although this 
representation does not allow to see a particular structural lattice, it allows to distinguish the 
local arrangement around each $i$ particle, and also how $\tilde{Q}_{6}$ is distributed in the 
jammed structure. Fig.~\ref{HeatmapPlots_SRp7321} (c)-(f) shows the distribution of $\tilde{Q}_{6}$
in the jammed packing for $\delta = 0.73$. Looking at Fig.~\ref{HeatmapPlots_SRp7321} (c) and 
Fig.~\ref{HeatmapPlots_SRp7321} (f), where large and small particles dominate the structure at 
different $X_{\mathrm{S}}$, one cannot see much difference between the distributions of $\tilde{Q}_{6}$.
This indicates that similar structures are obtained independently of $X_{\mathrm{S}}$, thus leading
to similar $\phi_J$, see Fig.~\ref{jamming}. For $\delta = 0.15$, the distribution 
of $\tilde{Q}_{6}$ shows a different scenario with $X_{\mathrm{S}}$. At $X_{\mathrm{S}} = 0.4$, 
the distribution of $\tilde{Q}_{6}$ is dominated by small particles, see Fig.~\ref{HeatmapPlots_SRp1548} (f),
while large ones do not contribute between $0.2 \leq \tilde{Q}_{6} \leq 0.8$, see 
Fig.~\ref{HeatmapPlots_SRp1548} (d). This happens due to the low number of large particles 
in the system, $N_{\mathrm{L}} \approx 32$. Thus, they tend to share only contacts 
with small particles. This leads to $\tilde{Q}_{6} < 0.2$ for large particles. For $X_{\mathrm{S}} = 0.1$ and 
second transition, small particles still dominate the distribution of $\tilde{Q}_{6}$ over large ones, see
Fig.~\ref{HeatmapPlots_SRp1548} (c, \,e). In this case, the distribution of $\tilde{Q}_{6}$ shows a structure 
where small particles form clusters of the same $\tilde{Q}_{6}$ value (specifically between 
$\tilde{Q}^{\mathrm{HCP}}_{6} = 0.48$ and $\tilde{Q}^{\mathrm{FCC}}_{6} = 0.57$). This indicates that 
small particles accumulate in HCP and FCC structures.

\begin{figure}[t]
 \centering \includegraphics[scale=0.3]{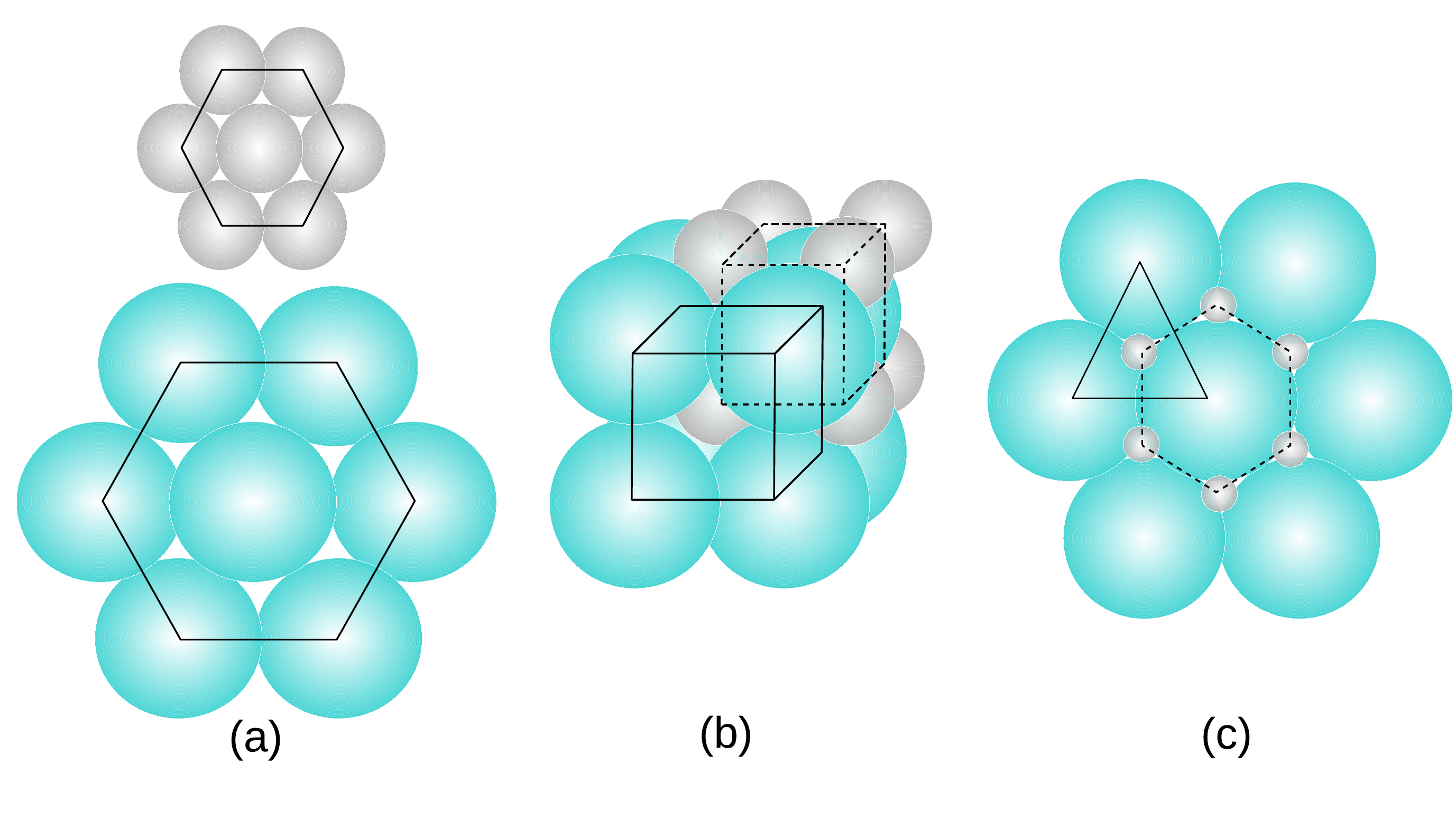}

 \caption{Illustration of typical structures found in bidisperse packings. (a) Hexagonal structures 
 are generally formed by LL and SS contact configurations. (b) For $\delta = 0.73$, the densest lattice is a
 cubic structure formed between LS (dashed line) and SL particle contacts (solid line). (c) For $\delta = 0.15$,
 triangular and hexagonal lattices are formed between SL (solid line) and LS particle contacts (dashed line) 
 to achieve the most efficient lattice.}
  \label{structures}
\end{figure}

The results presented above show that bidisperse packings not only form HCP and FCC crystal-like structures 
of the same particle size, but also other local structures are relevant in the packing. In addition to Large-Large 
and Small-Small particle contacts lead to specific local structures, contacts between large and small 
particles also tend to form completely different ones. Such local structures are already accounted 
for in the distribution of $P(\tilde{Q}_{6})$, but are difficult to distinguish from the six-fold lattice. 
To explore the different local structures, in the next section we will investigate how the individual 
structures of the contact configuration contribute to the jammed structure of the bidisperse packings.

\subsection{Partial $\tilde{Q}_{\ell}$ of local structures}

$\tilde{Q}_{\ell}$ is determined for each contact configuration: LL, SS, SL, and LS. For LL and SS, we determined $\tilde{Q}_{6}$ 
because they can form local hexagonal structures when are packed, see 
Fig.~\ref{structures} (a). Mixed contacts are treated differently. Large particles can be packed with 
small particles in different ways, depending on the size ratio, see Ref.~\cite{kumar2016tuning}. 
For $\delta = 0.73$, one can demonstrate that the most efficient lattice is a cubic lattice, 
where a small (large) particle with radius $r_{\mathrm{S}} = (\sqrt{3} -1) r_{\mathrm{L}}$ 
is in the center of a cube of large (small) particles, see Fig.~\ref{structures} (b). 
Thus, we calculate $\tilde{Q}_{8}$ for SL and LS, respectively. For $\delta = 0.15$, a small particle
with radius $r_{\mathrm{S}} = [(\sqrt{3} -1)/2\sqrt{3}] r_{\mathrm{L}}$ fits into a triangular
lattice of large particles, while a large particle fits into a hexagonal lattice of small ones, 
Fig.~\ref{structures} (c). Therefore, we use $\tilde{Q}_{3}$ for SL and $\tilde{Q}_{6}$ for LS. 
In this way, we get a better approximation of how each contact configuration assembles and contributes as the 
system approaches jamming, and also along the first and second jamming transitions.

Other $\tilde{Q}_{\ell}$ values can also be calculated on LL and SS contact configurations. 
For example, $\tilde{Q}_{4}$ and $\tilde{Q}_{8}$ can be determined to account for square 
and cubic local structures. However, we restrict ourselves to $\tilde{Q}_{6}$ since LL and SS
tend to form six-fold lattices, HCP and FCC, which are the densest lattices.
For the mixed contact configurations, $\tilde{Q}_{\ell}$ is determined by $\delta$ since it 
leads to the densest lattice formed between large and small particles, see Fig.~\ref{structures}.

\begin{figure}[t]

\centering \includegraphics[scale=0.24]{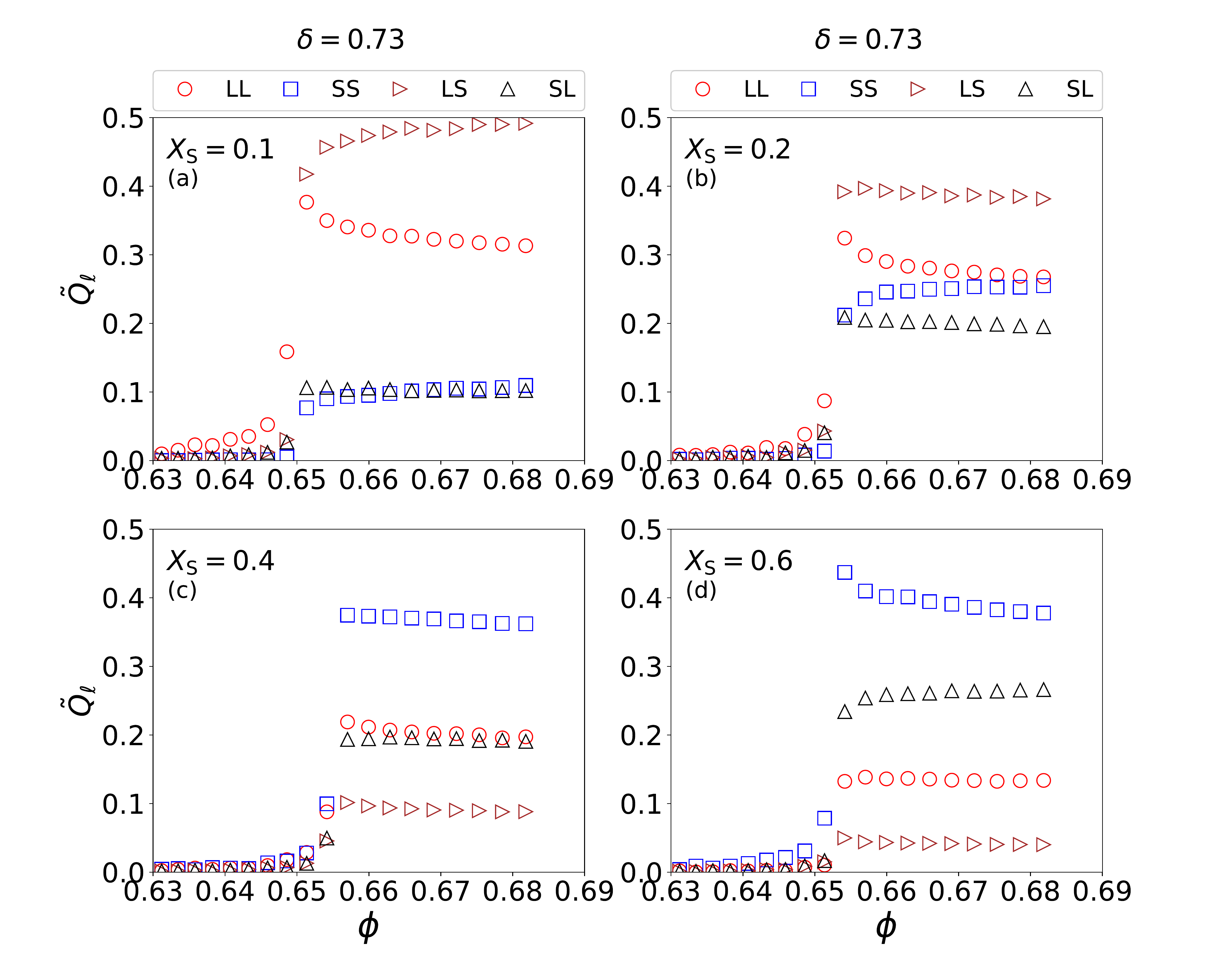}

 \caption{Partial $\tilde{Q}_{\ell}$ vs $\phi$ for $\delta = 0.73$ at different $X_{\mathrm{S}}$. 
 $\tilde{Q}_{6}$ is calculated for LL and SS contacts as they can form six-fold HCP and FCC 
 structures. While $\tilde{Q}_{8}$ is used for LS and SL contacts since a cubic structure is 
 the densest lattice. The jamming densities at each jump correspond to 
 $\phi_J(X_{\mathrm{S}} = 0.1) = 0.647$, $\phi_J(X_{\mathrm{S}} = 0.2) = 0.653$, 
 $\phi_J(X_{\mathrm{S}} = 0.4) = 0.655$ and $\phi_J(X_{\mathrm{S}} = 0.6) = 0.653$.}
  \label{q6_vs_phi_SRp7321}
\end{figure}

\begin{figure}[t]

\centering \includegraphics[scale=0.24]{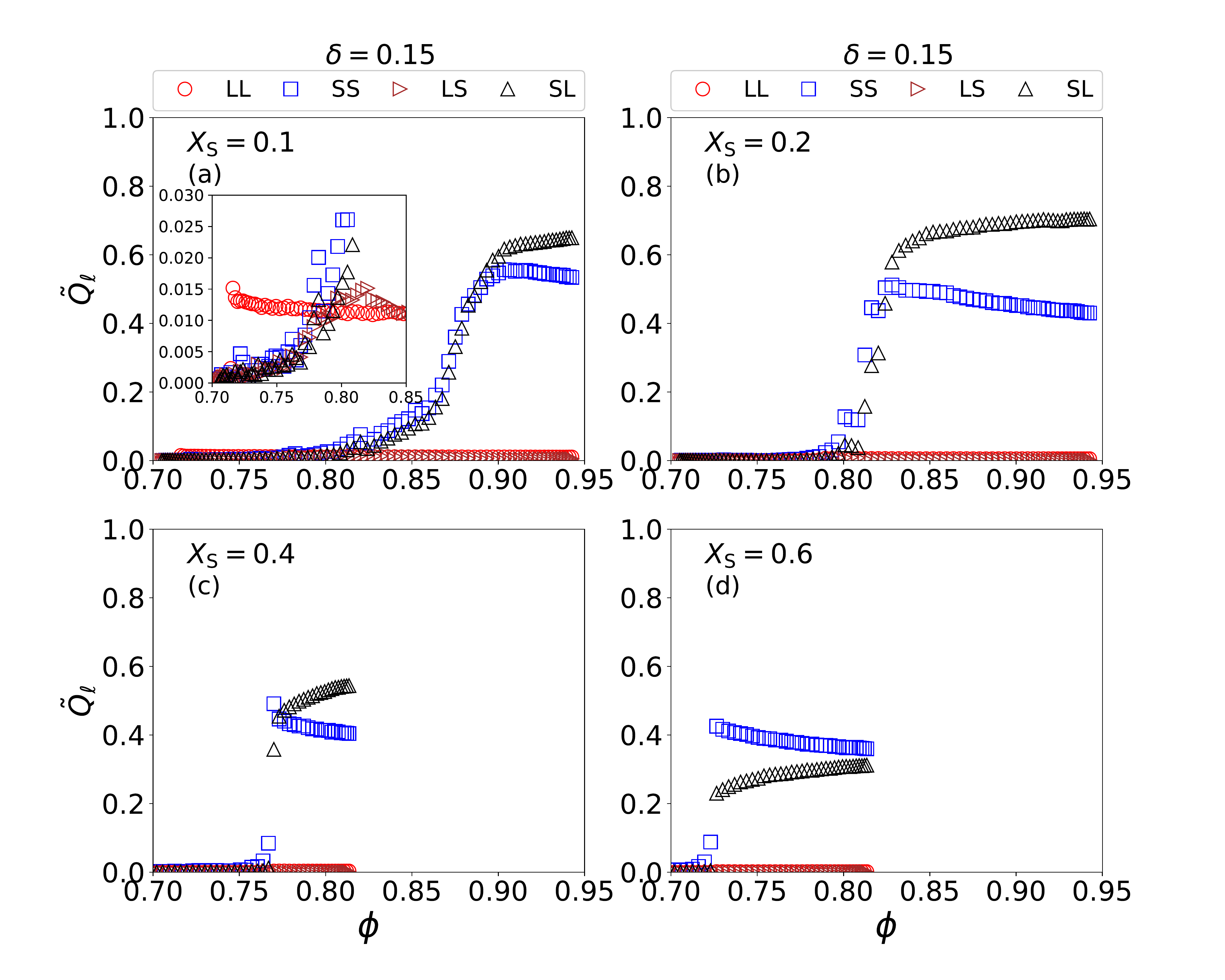}

 \caption{ Partial $\tilde{Q}_{\ell}$ vs $\phi$ for $\delta = 0.15$ at different 
 $X_{\mathrm{S}}$. $\tilde{Q}_{6}$ is calculated for LL and SS contacts as they can 
 form six-fold HCP and FCC structures. While $\tilde{Q}_{6}$ and $\tilde{Q}_{3}$ are
 used for LS and SL contacts, respectively. Each jamming density occurs at 
 $\phi_J^{\mathrm{L}}(X_{\mathrm{S}} = 0.1) = 0.717$ and 
 $\phi_J^{\mathrm{S}}(X_{\mathrm{S}} = 0.1) = 0.871$. 
 $\phi_J^{\mathrm{L}}(X_{\mathrm{S}} = 0.2) = 0.804$ and 
 $\phi_J^{\mathrm{S}}(X_{\mathrm{S}} = 0.2) = 0.820$. 
 $\phi_J(X_{\mathrm{S}} = 0.4) = 0.769$ and 
 $\phi_J(X_{\mathrm{S}} = 0.6) = 0.726$. 
 The superscripts $\mathrm{L}$ and $\mathrm{S}$ in $\phi_J$ represent the jamming of large and 
 small particles at different densities. The inset shows the rising point of
 each contact type contribution at low $\phi$.}
  \label{q6_vs_phi_SRp1548}
\end{figure}

Fig.~\ref{q6_vs_phi_SRp7321} shows the values of partial $\tilde{Q}^{\mathrm{ LL }}_{6}$, 
$\tilde{Q}^{\mathrm{ SS }}_{6}$, $\tilde{Q}^{\mathrm{ LS }}_{8}$ and $\tilde{Q}^{\mathrm{SL}}_{8}$ 
as a function of $\phi$ for $\delta = 0.73$ at different $X_{\mathrm{S}}$. We find that all partial 
$\tilde{Q}_{\ell}$ have similar density, namely $\phi_J \approx 0.65$, independent of $X_{\mathrm{S}}$. 
This agrees with the results shown in Fig.~\ref{fractpart_nu}, where the fraction of large, $n_{\mathrm{L}}$, 
and small particles, $n_{\mathrm{S}}$, show a jump at the same $\phi_J$. However, a clear difference 
in the structural arrangement of the contact configurations is observed when $X_{\mathrm{S}}$ changes. 
For $X_{\mathrm{S}} = 0.6$, $\tilde{Q}^{\mathrm{ SS }}_{6}$ dominates the jammed structure compared to 
$\tilde{Q}^{\mathrm{ LL }}_{6}$, indicating the formation of a fraction of six-fold structures in the 
system, see Fig.~\ref{q6_vs_phi_SRp7321} (d). $\tilde{Q}^{\mathrm{ SL }}_{8}$ dominates next the jammed 
structure over $\tilde{Q}^{\mathrm{ LS }}_{8}$ showing that there are higher local formations of cubic 
structures in which small particles are surrounded by large ones. When $X_{\mathrm{S}}$ decreases, the 
dominance of SS and SL in the jammed structure is completely exchanged by LL and LS, see Fig.~\ref{q6_vs_phi_SRp7321} (a). 
This change in structural dominance, indicated by $\tilde{Q}_{\ell}$, is due to the high number of 
large particles present at low $X_{\mathrm{S}}$ that tends to form local six-fold and cubic structures, 
see Fig.~\ref{structures} (b). One can speculate that there should be a certain value of $X_{\mathrm{S}}$
at which all partial $\tilde{Q}_{\ell}$ are equal. For $\delta = 0.73$, this could be at 
$X_{\mathrm{S}} = 0.28$, which corresponds to the 50:50 mixture, see Fig.~\ref{protocol}. 
However, this needs to be investigated and clarified in future work.

For $\delta = 0.15$ the scenario is different. $\tilde{Q}^{\mathrm{ SS }}_{6}$ 
and $\tilde{Q}^{\mathrm{ SL }}_{3}$ always dominate the jammed structure over $\tilde{Q}^{\mathrm{ LL }}_{6}$ 
and $\tilde{Q}^{\mathrm{ LS }}_{6}$ due to the low number of large particles.
We note that for $X_{\mathrm{S}} > 0.2$ the partial $\tilde{Q}_{\ell}$ 
jumps at the same $\phi_J$. This suggests that some HCP, FCC, and triangular structures are formed 
during jamming, which become more frequent as $\phi$ increases. For $X_{\mathrm{S}} < 0.2$, there is 
a structural decoupling in the packing depending on the contact configurations, becoming more evident 
for $X_{\mathrm{S}} = 0.1$. A first jump is observed at $\phi_J \approx 0.71$ which indicates the first 
jamming transition of the system. An amorphous structure with low formation of six-fold structures is obtained
with $\tilde{Q}^{\mathrm{ LL }}_{6} \approx 0.015$, see the inset in Fig.~\ref{q6_vs_phi_SRp1548} (a). 
Upon further compression, small particles begin to contribute to the jammed structure. This is indicated 
by the smooth increase in $\tilde{Q}^{\mathrm{ SS }}_{6}$, $\tilde{Q}^{\mathrm{ LS }}_{6}$, and 
$\tilde{Q}^{\mathrm{ SL }}_{3}$. Then, the structure is dominated by SS and SL contact configurations 
at $\phi \approx 0.77$. At the second transition, 
$\phi_J \approx 0.87$, where a large number of small particles suddenly jam, we find that 
$\tilde{Q}^{\mathrm{ SS }}_{6} \approx 0.28$, which is far from the typical values found in the local HCP and 
FCC packing structures. One can assume that the structure is disordered but endowed with a local 
order due to the existence of a fraction of HCP and FCC structures, see Fig.~\ref{HeatmapPlots_SRp1548} (a).

Fig.~\ref{q6_vs_phi_SRp7321} and Fig.~\ref{q6_vs_phi_SRp1548} display an unusual behavior of $\tilde{Q}_{\ell}$
for $\phi \gg \phi_J$: $\tilde{Q}_{\ell}$ decreases or increases with $\phi$ depending on the 
contact type. We think that this behavior is a consequence of over-compression of the system that generally 
leads to different particle penetrations and distortion of the packing structure. For a full discussion of the 
contact type penetration as a function of $\phi$, see Ref. \cite{petit2022bulk}.

\section{Summary and conclusion}
\label{SecVI}

We have studied the structural transition along the first and second jamming transition 
lines in bidisperse granular packings. The local contact orientational order (LCOR), 
$\tilde{Q}_{\ell}$, was introduced to quantify the local structures of the packings by 
considering particle contacts as nearest neighbors. This means that any sudden change in 
contacts will be reflected on $\tilde{Q}_{\ell}$. The global structure of each bidisperse 
packing was divided into contact configurations, i.e., Large-Large (LL), Small-Small (SS), 
Large-Small (LS), and Small-Large (SL), to quantify the role of each contact configuration 
in the jammed packing. For the LL and SS contact configurations, $\tilde{Q}_{6}$ was calculated 
with respect to the six-fold lattice, while LS and SL were treated differently because the 
local structures depend on $\delta$. For $\delta = 0.73$, $\tilde{Q}_{8}$ was determined for 
LS and SL because a cubic lattice is the most efficient one. For $\delta = 0.15$, 
$\tilde{Q}_{3}$ and $\tilde{Q}_{6}$ were calculated for SL and LS since they tend to form 
triangular and hexagonal lattices, respectively.

We found that for $\delta = 0.73$ all partial $\tilde{Q}_{\ell}$ show a sudden jump at the 
same $\phi_J$, but have different contributions as $X_{\mathrm{S}}$ changes. For example, at $X_{\mathrm{S}} = 0.1$, 
the jammed structure is dominated by a fraction of six-fold and cubic lattices of LL and LS, respectively. 
In contrast, for $X_{\mathrm{S}} = 0.6$, this scenario is dominated by the same local structures, 
but for SS and SL contact configurations. For $\delta = 0.15$, a different behavior is observed. 
For $X_{\mathrm{S}} = 0.1$, the contribution of each contact configuration to the jammed structure 
decouples as $\phi \to \phi_J$. At $\phi_J \approx 0.71$, $\tilde{Q}^{\mathrm{ LL }}_{6}$ shows a jump to 
a very low value, while the rest of the partial $\tilde{Q}_{\ell}$ are zero. This represents the 
formation of an amorphous jammed structure and thus the first jamming transition of the system. 
Upon further compression, we observe a structural change experienced by small particles,
marking the second jamming transition. The jammed structure at the second transition is 
dominated by a fraction of six-fold and triangular lattices of SS and SL contact configurations. 
For $X_{\mathrm{S}} > 0.2$, all $\tilde{Q}_{\ell}$ simultaneously jump to the same $\phi_J$, with 
the jammed structure dominated mainly by SS and SL. The results have shown that the 
evolution of both first and second jamming transitions is accompanied by a structural change 
in the system, where some HCP/FCC structures of large and small particles are formed.

The analysis of the structure factor has shown that the packing structures for $\delta = 0.73$ 
are similar showing with approximately the same $\phi_J$ as $X_{\mathrm{S}}$ changes. Although, 
each configuration type contributes differently to the total $S(q)$. Therefore, the 
jammed structures can be expected to be structurally similar for $\delta = 0.73$, 
regardless of the concentration of the small particles. Such an equivalence of 
packing structures needs to be confirmed and explained in future research.

We have also shown that the $S(q)$ immediately before jamming is identical to the $S(q)$ 
at jamming. This means that $S(q)$ cannot distinguish whether the system is jammed or not. 
Choosing a variable that responds to each transition is important for understanding how 
jammed structures arise and which contact configuration contributes the most. The introduction 
of the contact orientational order aims to shed light on the local structures
of the packings. Looking at the partial $\tilde{Q}_{\ell}$, one can determine how the contact 
configurations are packed when $\phi \to \phi_J$. This result could be crucial to understand 
how the strength of the bidisperse packing evolves. For example, it is of great importance
to know how the bulk modulus and other macroscopic properties of the packings depend on the packing 
structure. The structural features presented here are consistent with the 
jamming density and bulk modulus reported in Ref.~\cite{petit2022bulk} along the first and second 
jamming transition, which can suggest a possible connection between them. In particular,
the similar values of the jamming density and bulk modulus observed for $\delta = 0.73$ may be due 
to the formation of equivalent jammed structures independent of $X_{\mathrm{S}}$. For $\delta = 0.15$, 
the structures seem to correlate with the jamming density and bulk modulus, suggesting that 
the structure has an effect on them. Future work in this direction is ongoing.

\section*{Acknowledments}

We thank Thomas Voigtmann, Philip Born, and Michael A. Klatt for proofreading and fruitful discussions.
This work was supported by the German Academic Exchange Service (DAAD)
under grant n$^{\circ}$ 57424730.

\section*{Compliance with ethical standards}

Conflict of Interest: The authors declare that they have no conflict of interest.


%

\bibliographystyle{unsrt}
\bibliography{Ref}

\end{document}